\documentclass[usenatbib]{mn2e} 
\usepackage{epsfig}
\usepackage{subfigure}
\usepackage{amssymb}

\def\apj{ApJ}  \def\apjl{ApJL} 
\def\mnras{MNRAS} \def\aj{AJ}  \def\apss{Ap\&SS}
\def\nat{Nature} \def\pasj{PASJ}

\title{The dynamics of satellite disruption in cold dark matter haloes} 

\author[Choi, Weinberg, \& Katz]
{Jun-Hwan Choi\thanks{Current address : Department of Physics \& Astronomy, University of Nevada, Las Vegas, 4505 S. Maryland Pkwy, Las Vegas, NV, 89154-4002}\thanks{Email: jhchoi@physics.unlv.edu}, Martin D. Weinberg, Neal Katz
\vspace{0.3cm}\\
Department of Astronomy, University of Massachusetts, Amherst,  MA 01003}

\begin{document}

\maketitle

\begin{abstract}
  We investigate the physical mechanisms of tidal heating and
  satellite disruption in cold dark matter host haloes using N-body
  simulations based on cosmological initial conditions.  We show the
  importance of resonant shocks and resonant torques with the host
  halo to satellite heating.  A resonant shock (torque) couples the
  radial (tangential) motion of a satellite in its orbit to its phase
  space.  For a satellite on a circular orbit, an ILR-like resonance
  dominates the heating and this heating results in continuous
  satellite mass loss.  We estimate the requirements for simulations to
  achieve these dynamics using perturbation theory.  Both resonant
  shocks and resonant torques affect satellites on eccentric orbits.
  We demonstrate that satellite mass loss is an outside-in process in
  \emph{energy space}; a satellite's stars and gas are thus protected
  by their own halo against tidal stripping.  We simulate the
  evolution of a halo similar to the Large Magellanic Cloud (LMC) in
  our Galactic dark matter halo and conclude that the LMC stars have
  not yet been stripped.  Finally, we present a simple algorithm for
  estimating the evolution of satellite mass that includes both shock
  heating and resonant torques.
\end{abstract}

\begin{keywords}
  galaxies: evolution --- galaxies: interaction --- galaxies: haloes
  --- galaxies: kinematics and dynamics --- method: numerical --- 
  method: N-body simulation
\end{keywords}

\section{Introduction}
\label{sec:intro}

Physical processes affecting satellite galaxy evolution in their host
haloes are an important component of galaxy formation in the cold dark
matter (CDM) cosmogony as galaxies are
built up from the assembly of small structures.  This assembly includes
the process of satellite galaxies merging with their host
galaxies.  Moreover, recent CDM cosmological simulations predict the
existence of a large number of \emph{subhaloes} \citep{Ghigna98,
  Klypin99}. Consequently, understanding the detailed physical
processes affecting satellite evolution are key ingredients
to understanding galaxy formation in the CDM cosmogony.

Several basic questions about satellite halo evolution remain.  First,
how is the satellite stripped?  In other words, what parts of the
initial mass distribution might persist to the present day?  Second,
what is the rate of satellite halo disruption?  Third, how does a
satellite's internal structure evolve?  In a satellite-galaxy merger,
the stars and gas of the satellite galaxy can be stripped and become halo
stars and halo gas \citep[e.g.][]{QBB00,BKW01}.  The interaction
between the satellite galaxy and its host during the course
of the merger results in evolution of the satellite galaxy 
\citep[e.g.][]{MKL96}.  The remaining components of the satellite
galaxy merge with the host galaxy and this merging causes 
the host galaxy to gain mass \citep[e.g.][]{Murali02}.

Current cosmological simulations can only provide statistical
properties of subhaloes since even in the highest resolution
cosmological simulations \citep{Ghigna00, DLucia04, Diemand04, Gao04,
  OL04, Diemand.etal:07, Springel.etal:08} the detailed physical
processes of individual subhalo evolution have not been accurately
studied, owing to limited resolution.  To investigate the physical
processes in detail, we perform high resolution idealised simulations
with cosmologically motivated initial conditions instead of using
cosmological simulations.  In these idealised simulations, a live
satellite orbits in a static host halo.  Although too simplified to
reproduce a satellite's evolution in realistic detail, several authors
have used similar non-cosmological simulations to study satellite
disruption with alternative simulation methods
\citep[e.g.][]{Hayashi.etal:03,Kazantzidis.etal:04,Read.etal:06,
  Boylan-Kolchin.Ma:07,Penarrubia.Navarro.McConnachie:08}.  These
studies have demonstrated important features of the satellite
evolution such as mass loss history and density profile evolution.  In
this paper we use the higher resolution simulation to investigate
satellite disruption particularly focusing on the detailed physical
processes affecting satellite evolution such as resonant dynamics.

When a time-dependent force acts on a bound system such as a galaxy
or a dark matter halo, resonant interactions play an important role
in the system evolution.  Recent studies of resonant dynamics in
galaxy evolution claim that high resolution simulations are required
to accurately reproduce these resonant effects \citep{WK07a,WK07b} and 
\citet{WK07a} provide a procedure to determine minimum particle number
guidelines.  Since our idealised, high
resolution simulations are designed to satisfy these particle
number guidelines, they allow us to investigate the role of resonant
dynamics in satellite disruptions.

Resonant interactions couple a time-dependent perturbing force with
orbits in the system.  The frequency spectrum of the time-dependent
perturbing force characterises the interaction.  For a
satellite orbiting in its host halo, the satellite's orbital
frequencies, and possibly the rate of orbital decay, determine the
time-dependence of the external force.  A general, eccentric satellite
orbit in a spherical halo has both a radial and azimuthal frequency,
making the resonant coupling for an eccentric orbit complex.
Empirically, we may characterise the overall effects of the
interaction as a \emph{resonant shock} and a \emph{resonant torque}.
A resonant shock represents coupling with the radial orbital frequency
and a resonant torque represents coupling with the azimuthal orbital
frequency.

A resonant shock is a generalisation of the standard impulsive,
gravitational shock.  During a resonant shock, some orbits within the
satellite gain energy through resonant coupling even though they are
not in the impulsive limit, i.e. the time scale of the perturbation
near pericentre is much longer than the internal orbital time scale.
A resonant torque couples the rotation in the external potential to
orbits in the satellite.  For a resonant torque, the magnitude of the
external potential does not have to change; a change in the position
angle of the satellite frame relative to the centre of the halo is
sufficient to produce a torque.  Resonant shocks have been previously
considered and included in the impulsive approximation as an adiabatic
correction \citep[e.g.][]{Spitzer87,Weinberg94a,Weinberg94b,GO99}.
However, resonant torques have not been similarly considered in
satellite evolution studies although they have been extensively
investigated in the dynamics of barred galaxies.  In this study, we
will carefully investigate these resonant effects on satellite
evolution using this distinction.

Globular cluster evolution in a host galaxy is well established
\citep[e.g.][]{Spitzer87,Chernoff.Weinberg90}.  A globular cluster
experiences both tidal truncation and heating by both compressive
gravitational shocks and tidal shocks.  Because satellite halo
evolution in a host halo is similar to globular cluster evolution in a
host galaxy, many galaxy formation studies employ simple analytic
formulae taken from these globular cluster evolution studies to
estimate satellite galaxy evolution.  However, unlike globular
clusters, the satellite--host mass ratio is \emph{not} vanishingly
small.  This breaks the spatial symmetry in mass loss, as shown in
\citet{Choi.etal:07}, and changes the relative importance of resonant
coupling.

In this paper, we present numerical simulation results of satellite
galaxy disruption.  In \S\ref{sec:method}, we present an overview of
our numerical techniques: the N-body simulation code, the generation
of initial conditions, and the relevant perturbation theory.  In
\S\ref{sec:circ}, we present the results of a circular orbit
simulation.  We show that resonant torque effects result in
significant satellite mass loss.  In \S\ref{sec:ecc}, we present the
results of eccentric orbit simulations.  We show that satellite
heating by gravitational shocks at pericentre, which also includes
internal structure evolution, is the dominant process responsible for
disrupting the satellite.  In \S\ref{sec:stripping}, we show that the
process of satellite stripping is an outside-in process in satellite
\emph{energy} space.  Using this finding, we suggest an explanation
for the `missing' LMC stellar stream.  We also discuss the evolution
of the satellite density profile.  In \S\ref{sec:massloss}, we provide
an improved analytic estimate for satellite mass loss, and we
summarise in \S\ref{sec:summary}.

\section{Numerical methods}
\label{sec:method}

Our N-body simulations use a three-dimensional self-consistent field
code \citep[SCF, also known as an expansion
code;][]{cbrock72,cbrock73,ho92,Weinberg99}.  The expansion code
calculates bi-orthogonal basis sets of density-potential pairs and
computes the gravitational potential of the system using these basis
sets.  There are two reasons that an expansion code is an appropriate
potential solver for our study.  First, the expansion basis can be
chosen to follow the structure over an interesting range of scales and
simultaneously suppress small-scale noise.  In contrast, the noise
from two-body scattering can arise at all scales in direct-summation,
tree, and mesh based codes.  Small-scale noise can also give rise to
diffusion in conserved quantities, which can then lead to unphysical
outcomes particularly for studies of long-term galaxy evolution
\citep{WK07a,WK07b}.  Second, the expansion code is computationally
efficient; the computational time increases only linearly with
particle number and with modest overhead.  Hence, the expansion code
permits the use of a much larger number of particles than most other
codes for the same computational cost.

The expansion code is not adaptive.  The largest efficiency obtains
when the basis resembles the galaxy.  Otherwise, the expansion code
requires a large number of basis function pairs, which introduces
small-scale noise and results in a greater computational overhead.
These constraints have been minimised by some recent improvements
\citep{Weinberg99,Choi.etal:07}.  First, employing a numerical
solution of the Sturm-Liouville equation, the initial galaxy model can
be used as the zeroth-order basis function for the expansion code.
Then, the expansion code requires only a modest number terms to
accurately compute the potential.  Second, we accurately trace the
density--potential centre of the expansion during the course of a
simulation \citep[see][for details]{Choi.etal:07}.  Third, to reduce
truncation error, we separately track the centre of mass motion of the
satellite and the relative motion of the satellite centre.  With these
improvements, the expansion simulations are now able to achieve a
sufficiently high central resolution with sufficiently low noise to
allow the investigation of the detailed dynamical processes important
for satellite disruption.

All the halo models used in our simulations are based on the
cosmologically-motivated universal CDM halo \citep[][hereafter
NFW]{NFW97}, $\rho(r) \propto r^{-1}(r+r_{s})^{-2}$, where $r_{s}$ is
a scale length characterised by the concentration parameter
$c=R_{vir}/r_{s}$ and $R_{vir}$ is the virial radius of the halo.  We
represent the host halo potential by a concentration $c=15$, static
NFW halo.  The NFW profile has infinite extent but a real satellite is
tidally truncated.  Furthermore, the host halo tidal field will affect
a satellite halo even before the satellite halo passes within the host
halo's virial radius, but it is computationally expensive to simulate
the evolution of a satellite from such large radii.  Since the
objective of our study is only to understand the physical processes
responsible for satellite disruption, it is not necessary to simulate
the satellite's entire evolutionary history.  Therefore, we begin the
simulation with the satellite in the host halo on the desired orbit
and include the host halo tidal field when we generate the satellite's
initial conditions.  

Our simulations ignore the effects of dynamical friction and the
subsequent reaction of the host halo since the satellite masses of
interest are often much smaller than that of the host halo. According
to \citep{Boylan-Kolchin.Ma:07}, when the satellite-to-host mass ratio
is smaller than 0.1, the dynamical friction hardly makes the satellite
decay within half a Hubble time.  The gravitational back-reaction of
the host halo has a complex structure with an amplitude proportional
to mass ratio.  In many cases, the halo reaction is dominated by
barycentric motion of the halo center induced by the satellite
perturbation.  We will quantitatively discuss these dynamics in a
later paper.  Since our main objective in this paper is an understanding
of the physical processes of satellite disruption, we do not expect
omission of the halo response to affect our conclusions.

\begin{figure}
  \centerline{\includegraphics[width=1.0\columnwidth,angle=0]
    {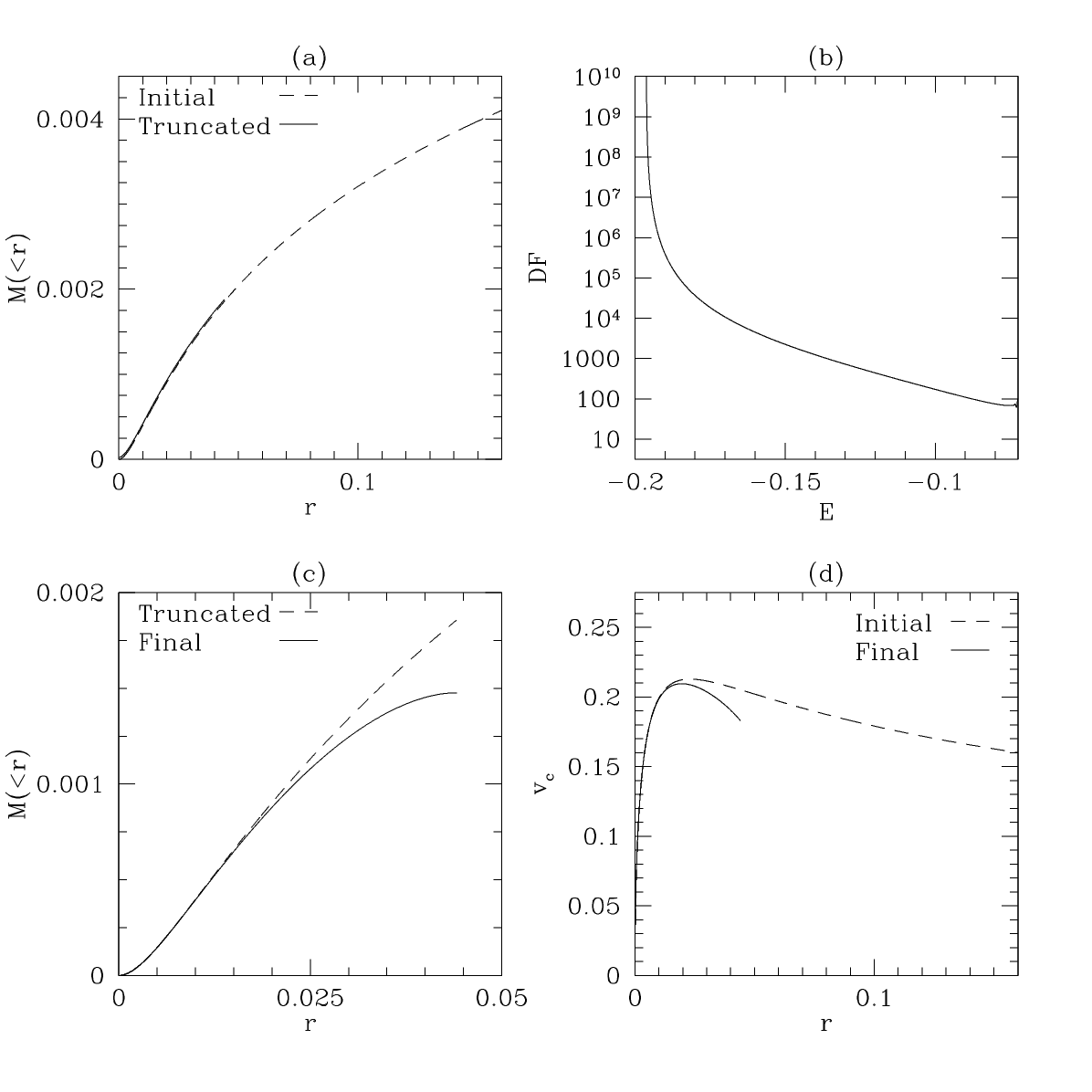}}
  \caption{The effect of our truncation procedure on a satellite's
    initial NFW profile.  (a) The density profiles of the initial NFW
    model and the truncated model.  (b) The distribution function
    computed using Eddington inversion.  (c) The density profiles of
    the truncated model and the final satellite model.  (d) The
    circular velocity profiles of the initial NFW model and the final
    satellite model.  The truncation radius in this figure is $x_{e}$
    and the satellite is on a circular orbit at $R=0.4R_{vir}$.}
  \label{fig:procedure_SateF}
\end{figure}

The details of the procedure that we use to generate a satellite's
initial conditions are as follows.  We start with an initial NFW
satellite halo model.  Since the maximum circular velocity is a better
measure of a satellite's initial size than its mass, owing to ongoing
mass loss, we characterise the satellite's size by its circular
velocity $V=\sqrt{M_{vir}/R_{vir}}$.  We choose the initial satellite
halo model such that the circular velocity of the satellite $V_{sat}
=\frac{1}{6} V_{host}$, which means that the initial virial mass of
the satellite is 0.0046.  Throughout this paper we use \emph{system}
units unless otherwise specified with $G=1$, $M_{vir, host}=1$, and
$R_{vir, host}=1$.  If scaled to the Milky Way, this satellite halo
corresponds to the size of the Sagittarius dwarf dark matter halo
\citep{Majewski04}.  We then determine the truncation radius, for
which we discuss several choices below, assuming that the satellite is
on a circular orbit with $R=0.4R_{host, vir}$.  We call this fixed
distance the tidal distance.  As shown in Fig.
\ref{fig:procedure_SateF}a, the truncated model is identical to the
initial NFW model but is chopped at the truncation radius.  After
truncating the initial satellite model at this radius, we perform an
Eddington inversion to compute the corresponding distribution function
\citep[][Chapter 4]{BT87}, as shown in Fig.
\ref{fig:procedure_SateF}b.  We calculate the final satellite density
profile by integrating this distribution function over velocity.  In
contrast to the truncated model, the final satellite model does not
show a sudden drop at the outer edge of its density profile (see
Fig. \ref{fig:procedure_SateF}c).  Owing to this smooth outer profile,
the final satellite model is closer to equilibrium than the truncated
model.  We use this final satellite model as our initial satellite
halo model.  Fig. \ref{fig:procedure_SateF}d shows the circular
velocity profiles of the initial and final satellite model; even
though the final satellite model is considerably truncated compared to
the initial model and has lost 2/3 of its initial mass, the maximum
circular velocities of both models are nearly the same.  We realise
the satellite phase space using an acceptance-rejection method.  All
realisations are assumed to have an isotropic velocity distribution
and are made of $10^{6}$ equal-mass particles.

\begin{figure}
  \centerline{\includegraphics[width=1.0\columnwidth,angle=0]
    {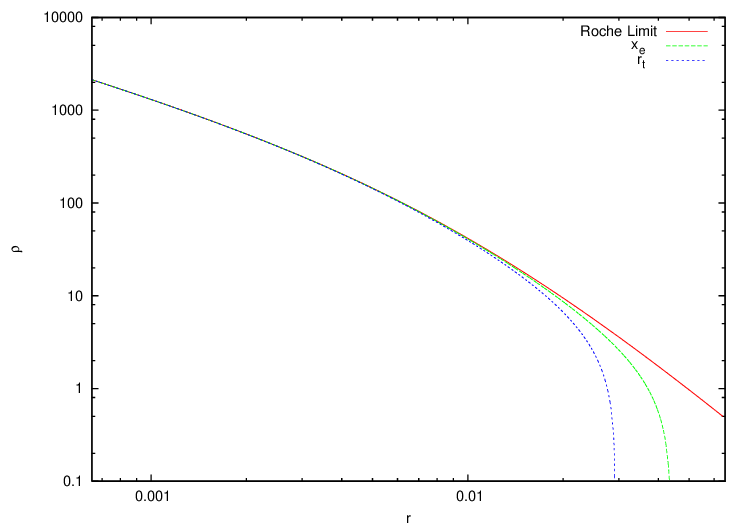}}
  \caption {Density profiles for three satellite initial conditions
    truncated at three different radii: the Roche limit (red), $x_{e}$
    (green), and $r_{t}$ (blue).}
  \label{fig:models}
\end{figure}

The Roche limit, the critical-point radius $x_{e}$, and the tidal
cut-off radius ($r_{t}$) \citep[][Chapter 5]{Spitzer87} are all
natural choices for the truncation radius.  The Roche limit is the
radius where the satellite density and the host halo density are the
same.  The critical-point radius, $x_{e}$, defines the unstable
zero-velocity equilibrium; that is, the unstable equilibrium in the
effective potential $d\Phi_{eff} = 0$.  For ease in computing initial
conditions and idealised tests in later sections, we define a
\emph{sphericised} centrifugal potential term of the form
\[\frac12\Omega^2\alpha r^2 \] 
with $\alpha\in[0,1]$ (see Appendix \ref{sec:pot_eff} for details). 
For initial conditions, we set $\alpha=1$ to make the
satellite smaller and hence to minimise satellite mass loss owing to
initial adjustments when it is first placed in the external
potential. The tidal cut-off\footnote{We follow Spitzer (1987) in
  defining the tidal cut-off radius as that derived for point-mass
  potentials with vanishingly small satellite masses to avoid any
  further confusion in terminology.  For extended gravitational
  potentials, the numerical value will differ from 2/3.}  is the
radius where $r_{t}=2x_{e}/3$.  This is the radius of the critical
potential surface perpendicular to the direction toward the host halo
centre so it is the maximum radius that a spherical satellite halo can
have and not extend beyond the critical potential surface in any
direction.  We plot the density profiles of the truncated satellite
halo models in Fig. \ref{fig:models}.  To obtain the final density
profiles for the satellite halo models truncated at $x_{e}$ and
$r_{t}$ we apply the Eddington inversion procedure. However, we do not
apply this procedure when we truncate at the Roche limit since it
would change the outer density, making it differ from that of the host
halo, and it would no longer be the Roche limit.  Fig.
\ref{fig:models} shows that at the given tidal distance of $0.4$, the
satellite truncated at the Roche limit is the largest and the
satellite truncated at $r_{t}$ is the smallest.

\begin{figure}
  \centerline{\includegraphics[width=1.0\columnwidth,angle=0]
    {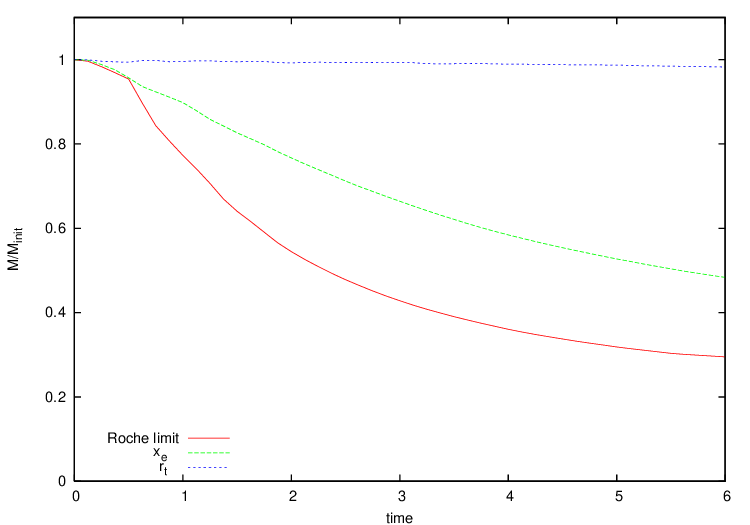}}
  \caption {The evolution of fractional mass with time for satellites
    on circular orbits.  Curves show the mass fractions remaining for
    satellites truncated at the Roche limit (red), $x_{e}$ (green),
    and $r_{t}$ (blue).  All three simulations exhibit continuous mass
    loss.  The satellite truncated at the Roche limit loses the most mass
    and the satellite truncated at $r_{t}$ loses the least mass.}
  \label{fig:massloss.Circ}
\end{figure}

Fig. \ref{fig:massloss.Circ} shows the evolution of mass from circular orbit
simulations for the three satellites shown in Fig. \ref{fig:models}
(see \S\ref{sec:circ} for a detailed discussion).  All three
satellites show continuous mass loss.  The amount of mass loss
correlates with the satellite size: the satellite truncated at the
Roche limit loses the largest mass fraction and the satellite
truncated at $r_{t}$ loses the least mass.  We choose $x_e$ for our
fiducial truncation radius on physical grounds. $x_e$ is the
transition point beyond which satellite particles on a circular orbit
become unbound. At the Roche limit radius for an NFW halo, the host
halo potential dominates the gravitational potential of the satellite.
In contrast, using the tidal cut-off radius, $r_{t}$, is a sensible
choice for globular clusters, which have already orbited many times
around a galaxy and have already been severely truncated, but using
$r_t$ could underestimate a satellite halo's size since it has made
only a few complete orbits.  Hence, we restrict our study to initial
satellite haloes that are truncated at $x_{e}$.

As mentioned in \S\ref{sec:intro}, resonant interactions are expected
to dominate the tidal heating.  To accurately reproduce these resonant
interactions, N-body simulations need to satisfy several numerical
criteria.  \citet{WK07a} proposed explicit requirements for these
criteria.  First, a sufficient number of particles are required to
cover the phase space near resonance (hereafter, the \emph{coverage}
criterion).  Second, a sufficient number of particles are required to
reduce artificial diffusion.  Artificial diffusion can come from both
the gravitational forces of individual particles (hereafter, the
\emph{small-scale noise} criterion) and the potential fluctuations caused
by Poisson noise (hereafter, the \emph{large-scale noise} criterion).
Besides these particle number criteria, the potential solver must also
be able to resolve the scale of the resonant potential and the
realised phase space distribution must cover this region.  In our
study, we will verify that our simulations satisfy all of these
criteria.

Although a simulation can correctly reproduce resonant interactions,
it is hard to provide a detailed accounting of the individual
resonances.  For intuitive guidance, the resonant interaction effects
can be investigated using perturbation theory. We use a numerical
perturbation theory calculation as in \citet{WK07a} to investigate
resonant interaction effects.  In this approach, one begins with the
numerical integration of the perturbed orbit-averaged Hamilton
equations in a fixed potential for the entire phase space.  This step
may be followed by an update of the gravitational potential.  Since
this perturbation calculation uses the same satellite halo
realisation, comparison with the N-body simulation result is
straightforward.  A comparison between the results of the N-body
simulation and those of the numerical perturbation theory calculation
provides us with strong evidence for the existence of resonant
dynamics and a definite understanding of its role in satellite
evolution.

\section{Satellite disruption on a circular orbit}
\label{sec:circ}

To study the resonant torque effect in isolation, we simulate a
satellite on a circular orbit, which eliminates the gravitational
shock.  However, as long as the satellite's rotation does not match
its orbital rotation, i.e. it is not \emph{tidally locked}, the
satellite will experience a time-dependent azimuthal force variation
that can result in a net resonant torque.

\subsection{Investigation of the resonance torque effect}
\label{sec:res_torque}

As described in Fig. \ref{fig:massloss.Circ}, satellites on a
circular orbits continuously lose mass and have lost a significant
fraction of their original mass at the end of the simulation
(approximately the age of the Universe).  If a transient adjustment of
the satellite halo to its host halo potential were the cause of the
mass loss, the mass loss should lessen and stop in several dynamical
times as the satellite adjusts to the halo potential.  Therefore,
continuous mass loss from truncated satellite haloes must result from
satellite heating and not from a transient initial adjustment.  In a
circular orbit simulation, the magnitude of the external potential
does not vary and, therefore, there is no gravitational shock.  The
only mechanism that can disrupt a satellite under these circumstances
is a resonant torque.

The dynamics of the resonant torque are similar to the dynamics of the
bar-halo interaction.  Owing to the coupling with the time-dependent
perturbation, the orbits whose frequencies are nearly commensurate
with the perturber's frequency are torqued.  The commensurability
condition is
\begin{eqnarray}
  l_{1}\Omega_{1} + l_{2}\Omega_{2} = l_{3}\Omega_{pert}
  \label{eq:res}
\end{eqnarray}
where $\Omega_{1}$ and $\Omega_{2}$ are the radial and azimuthal
frequencies of a satellite's orbit, $\Omega_{pert}$ is the perturber's
frequency, and $l_{1}$, $l_{2}$, and $l_{3}$ are integers.  The orbits
that satisfy equation (\ref{eq:res}) are called \emph{resonant} orbits
and define loci in phase space for each triple of integers
$l_1:l_2:l_3$.  The resonant torque heats the satellite, reduces the
satellite's binding energy, and enhances its mass loss.  Although
particular resonant orbits receive or lose angular momentum, mass loss
is not confined to the resonant orbits.  Since a satellite is a
self-consistent gravitational system, the work required to apply the
torque reduces the entire satellite's binding energy and unbinds the
most weakly bound halo material.  Satellite mass loss in general will
be discussed in \S\ref{sec:stripping}.

Perturbation theory calculations provide helpful physical insights
into the nature of these resonant dynamics.  The angular momentum
transfer by the resonant interactions may be approximated by a
second-order, time-dependent perturbation calculation
\citep{Weinberg04,WK07a,WK07b}.  These calculations become
prohibitively complicated for real astronomical systems with
multiple time scales. First, owing to the finite age of the galaxy and
the time-dependence of the perturbation, the frequency spectrum
becomes broader.  Secondly, some resonances need a longer time period
than the galaxy lifetime to converge into the time-asymptotic limit
and, in the interim, the first-order transient features may strongly
affect the response.  More detailed discussions of perturbation theory
and resonant dynamics are presented elsewhere
\citep{TW84,Weinberg04,WK07a,WK07b}.  For our calculations here, we
represent the host halo potential by an orbiting perturber around a
stationary satellite with the distance of the perturber to the
satellite equal to the distance between the satellite and the host
halo centre in the simulation.  The mass of the perturber is the
enclosed mass of the host halo inside the satellite's orbit in the
simulation, and the perturber's frequency is the same as the
satellite's orbital frequency.  We investigate the resonant torque
using the following three steps: (1) we find the resonances that are
located within the satellite, (2) we estimate the particle number
requirements for a given resonance using the criteria from
\citet{WK07a}, and (3) we compute the individual resonant effects
using a numerical perturbation theory calculation and compare them to
the simulations.

\begin{figure}
  \centerline{\includegraphics[width=0.9\columnwidth,angle=0]
    {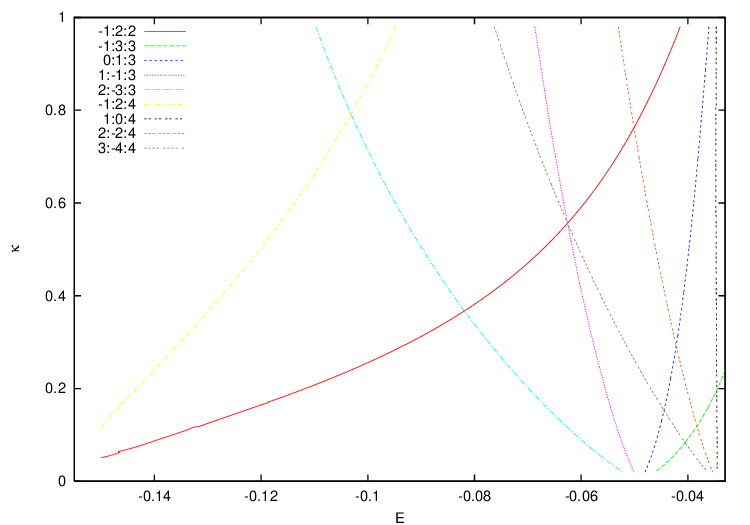}}
  \caption{The locations of resonances within the satellite in
    energy-angular momentum space.  The low-order resonances with $-10
    \le l_{1} \le 10$ and $l=1,2,3,4$ are shown (see text for
    details). }
  \label{fig:FindRes_XT}
\end{figure}

Fig. \ref{fig:FindRes_XT} shows the locations of the resonances
within the satellite phase space.  The phase space location is
represented by energy ($E$) and normalised angular
momentum\footnote{The quantity $\kappa\equiv J/J_{max}(E)$ where $J$
  is a particle's angular momentum and $J_{max}(E)$ is the maximum
  angular momentum at a given energy.} ($\kappa$).  The range of
resonances we examine is $-10 \le l_{1} \le 10$ and we restrict
ourselves to $l=1,2,3,4$, where $l$ is the spherical harmonic of the
perturber, in this case the primary galaxy.  Owing to symmetry,
$l_{2}$ and $|l_{3}|$ should be equal to or less than $l$.  Since
$l_{2}$ and $l_{3}$ are indices of the azimuthal expansion in
spherical harmonics for the halo and perturber respectively, $l_{2}$
and $l_{3}$ should have the same parity as $l$ \citep{TW84}.  As one
can see in Fig. \ref{fig:FindRes_XT}, nine resonances are located
within the satellite.

\begin{figure}
  \centerline{\includegraphics[width=0.9\columnwidth,angle=0]
    {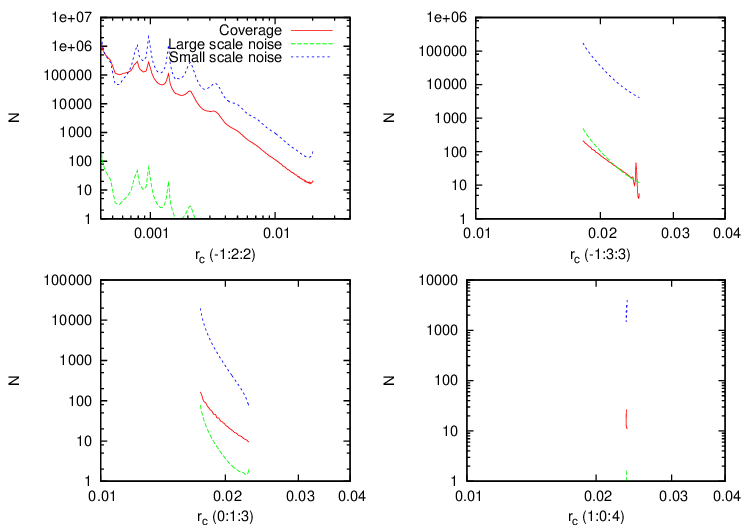}}
  \caption {Particle number requirements for the -1:2:2, -1:3:3,
    0:1:3, and 1:0:4 resonances.  These are the strongest four
    resonances for a satellite on a circular orbit.  Each panel shows
    the three particle number requirements from \citet{WK07a} for
    equal mass particles within the initial truncated satellite. They
    are plotted versus $r_c$, the radius of a circular orbit of a
    given energy in units of the host's virial radius.}
  \label{fig:Nreq_XT}
\end{figure}

The required numbers of equal mass particles within our truncated
satellite for the four strongest resonances are presented in Fig.
\ref{fig:Nreq_XT}.  There are three particle number criteria for each
resonance: coverage, small-scale noise, and large-scale noise
\citep{WK07a}.  According to this estimate, we require more than
$10^{5}$ satellite particles, which translates to more than $3\times 10^5$
within the initial virial radius, to correctly reproduce the -1:2:2
resonance. This resonance requires the largest number of particles and is
also the most important resonance as we show below.  Hence, our
$10^{6}$ equal-mass particle halo simulation easily satisfies the
necessary criteria (see Fig. \ref{fig:Nreq_XT}) for all the
important resonances.  The small-scale noise criterion is not relevant
for our expansion-code simulations.  However, if the N-body simulations
were to suffer from small-scale noise, such as the case in N-body
simulations using direct-summation, trees, or meshes, an order of
magnitude larger particle number would be required for these
resonances to be modelled correctly.

\begin{figure}
  \centerline{\includegraphics[width=0.9\columnwidth,angle=0]
    {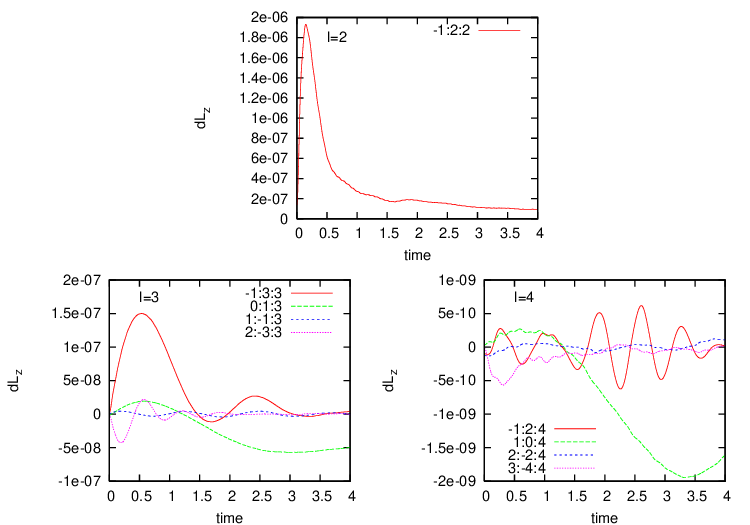}}
  \caption{The angular momentum transferred from the perturber to the
    satellite as function of time.  The panels show the $l=2, 3, 4$
    contributions as labelled. The -1:2:2 resonance (top panel)
    dominates the torque.}
  \label{fig:ResPot_XT}
\end{figure}

Fig. \ref{fig:ResPot_XT} shows the amount of angular momentum
deposited by an orbiting perturber calculated using numerical
perturbation theory for different resonances. The most significant angular
momentum change is
mediated by the -1:2:2 resonance.  The -1:3:3 and 0:1:3 resonances are
the strongest among the $l_{3}=3$ resonances, and the 1:0:4 resonance
is the strongest resonance among the $l_{3}=4$ resonances.  The amount
of angular momentum deposited by the -1:2:2 resonance is more than an
order of magnitude larger than the amount of angular momentum
deposited by any of the $l_{3}=3$ resonances and three orders of
magnitude larger than the amount of angular momentum deposited by any
of the $l_{3}=4$ resonances.  Hence, the -1:2:2 resonance 
dominates the resonant satellite torque.

\begin{figure*}
  \centerline{\includegraphics[width=0.9\textwidth,angle=0]
    {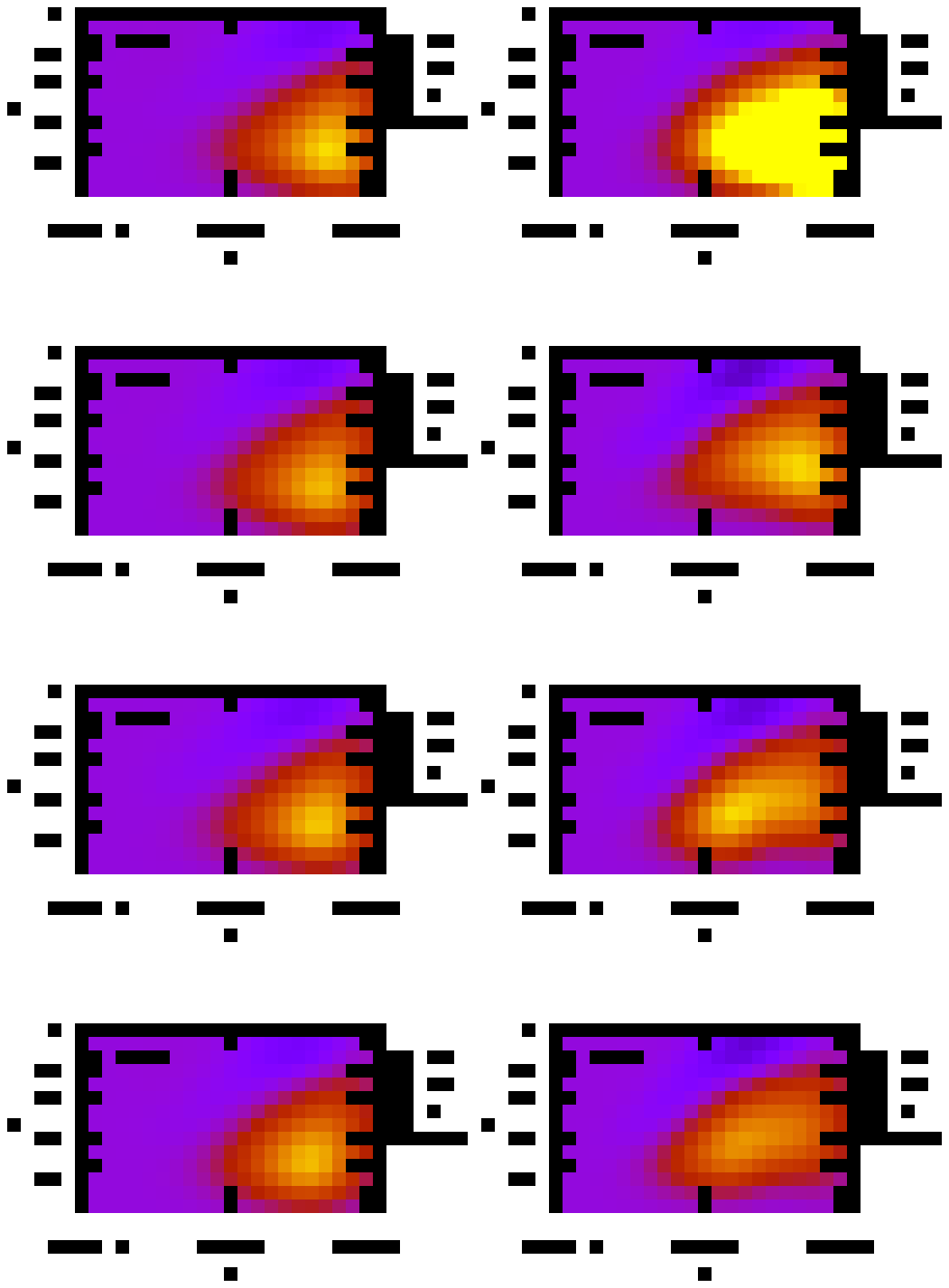}}
  \caption { The distribution of the relative change in $L_{z}$
    in phase space for the N-body simulation (left panels) and the
    numerical perturbation theory calculation (right panels) The
    perturbation theory calculation includes all the resonances in Fig.
    \ref{fig:FindRes_XT}. Except for $T=0.0 \;- \;1.0$, the results
    show good general agreement.  This agreement provides robust
    evidence for resonant effects in the circular orbit simulation.}
  \label{fig:plotdk}
\end{figure*}

Fig. \ref{fig:plotdk} shows the location of the angular momentum
transferred through resonant interactions in phase space by plotting the
distribution of the change in $L_{z}$ in phase space, $\Delta L_{z}$,
at different times from both the N-body simulation and the numerical
perturbation theory calculation. We normalise this change by the total
angular momentum at each energy $E$ since $L(E)$ at fixed $\kappa$
increases with energy, which biases the absolute $\Delta L_{z}$
distribution toward higher energies.  By normalising to the total
angular momentum at a given energy bin, the \emph{relative} $\Delta
L_{z}$ distribution enhances $\Delta L_{z}$ features at low energy.
Each panel shows the relative $\Delta L_{z}$ over a fixed time span.
Both the phase space location and magnitude of the angular momentum
change from the simulation agrees well with the numerical perturbation
calculation.  The perturbation theory only includes the resonant
torque and therefore this agreement is strong evidence that torque by
resonant interactions is the major mechanism responsible for circular
orbit satellite evolution.

Nonetheless, some minor discrepancies remain.  First, the amplitude of
the relative $\Delta L_{z}$ from the perturbation calculation is
larger than that from the simulation at $T=$0.0--1.0.  This results
from the abrupt introduction of the perturbation. 
In our calculations, we abruptly introduce the external potential 
(for the simulation) and perturbations by resonances (for the 
perturbation calculation) to the initial satellite. This abrupt 
introduction may induce a readjustment of satellite halo equilibrium and/or 
cause non-linear features, which could cause the discrepancy.
Second, the shape of the region of angular momentum
change is mildly different in the simulation compared to the
perturbation calculation.  In particular, the simulation shows a
relative $\Delta L_{z}$ distribution in the lower right region of
phase space, while the perturbation calculation does not.  The phase
space responsible for this difference in the $\Delta L_z$ distribution
has low binding energy and is easily stripped.  The simulation results
in Fig. \ref{fig:plotdk} shows the distribution of the relative
$\Delta L_{z}$ for only the unstripped particles.  Unlike in the
simulations, the satellite in the perturbation calculation does not
lose mass.  Hence, the few minor disagreements between the simulation
and the perturbation theory are a natural consequence of the
idealisation required to compute the perturbation theory and does not
invalidate our primary conclusion: resonant torque is the major
mechanism responsible for satellite disruption for a satellite on a
circular orbit.

The absence of a $l=1$ resonance contribution is a consequence of
satellite truncation.  Roughly, the tidal radius of the satellite
coincides with the corotation radius of the satellite since
$\Omega_{pert} \sim \Omega_{2}$ at $r_t$.  The satellite halo is
isotropic so $| \frac{\Omega_{pert}}{\Omega_{2}} | < 1$ and it follows
from equation (\ref{eq:res}) that $|
l_{1}\frac{\Omega_{1}}{\Omega_{2}}+l_{2} | < 1$ for a resonance within
the satellite.  Since $\Omega_{1}/\Omega_{2}$ is always larger than 1,
$|l_{1}\frac{\Omega_{1}}{\Omega_{2}}+l_{2} |$ has its lowest value
when $l_{1}=1$ and $l_{2}=-1$, making the location of the 1:-1:1 resonance
outside of the satellite.  Hence, if the satellite is tidally
truncated there is no $l=1$ resonance contribution to satellite
disruption.

\subsection{Verifying mass loss by the resonant torque}
\label{sec:ml_torque}

The simulation results and the perturbation theory calculation agree
on the amplitude and phase space location of the angular momentum
exchange, which confirms the importance of the resonant torque.
However, there is another process that may cause mass loss:
re-equilibration of the satellite after some particles are tidally
stripped by the host halo potential.  This readjustment in the profile
forces some particles to move beyond the tidal radius and also
results in continuous mass loss.  In our simulation, the satellite
initial conditions are designed to reduce the mass lost through this
process by truncating satellites at the tidal radius but, owing to the
Eddington inversion, a small amount of mass loss is inevitable.  To
elucidate the affect of resonant torque on satellite mass loss, we
conduct two types of idealised numerical experiments.  In the first,
we add angular momentum to the simulation using the torque determined
from perturbation theory, tidal truncation, and the centrifugal and
Coriolis forces.  We call these tests the \emph{synthetic force}
experiments.  In the second, we simulate satellites corotating with
their orbital revolution, i.e. as if there were tidally locked.  We
call these the \emph{rotating satellite} experiments.

\begin{figure}
  \centerline{\includegraphics[width=1.0\columnwidth,angle=0]
    {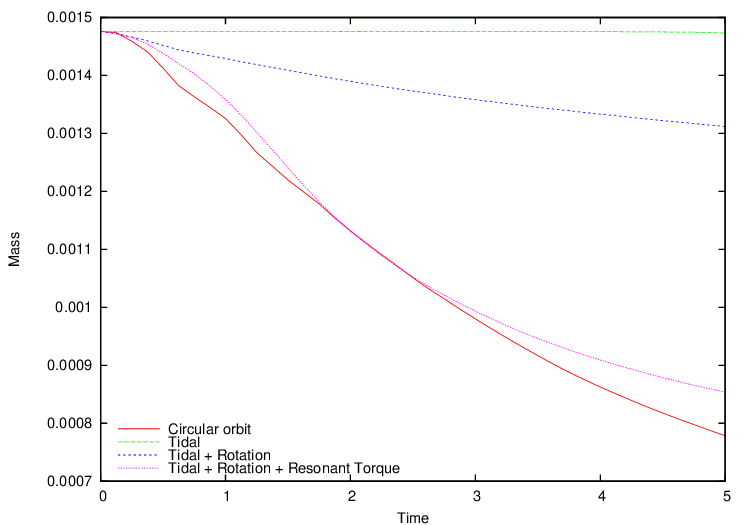}}
  \caption{The mass loss histories from the synthetic force
    experiments (see text).}
  \label{fig:massloss.iso}
\end{figure}

The synthetic force experiments estimate satellite mass loss by adding
known mass loss mechanisms to a non-orbiting satellite simulated in
isolation.  Satellite particles outside the tidal radius do not feel
the satellite's gravity, allowing particles outside this radius to
escape easily. As discussed in \S\ref{sec:method}, we take $x_e$ to be
the tidal radius, not $r_t$.  Satellite mass loss can drive structural
evolution and change $\Phi_{sat}(r)$, hence we update $x_{e}$ during
the course of the simulations.

We present the results of our synthetic force experiments in Fig.
\ref{fig:massloss.iso}.  The curve labelled \emph{Tidal} (long-dashed)
in Fig.\ref{fig:massloss.iso} only includes tidal truncation and
assumes $\alpha=0$ in the tidal radius calculation (see
\S\ref{sec:method} and Appendix \ref{sec:pot_eff}).  We include the
centrifugal force effect by including it when calculating $x_e$ and we
now set $\alpha=0.75$ \footnote{Note that $\alpha$ is a free parameter
  and can be between $0$ to $1$.}.  We include the Coriolis force by
adding it to the force used to update each particle's velocity at
every time step.  The mass evolution of the resulting simulation is
labelled \emph{Tidal+Rotation} in Fig.\ref{fig:massloss.iso}
(short-dashed) .  Finally, we include resonant torques by adding
angular momentum at the rate calculated from the numerical
perturbation theory calculation plotted in Fig.\ref{fig:plotdk}
between $T=1.0$ and $T=2.0$ according to each particle's energy and
angular momentum.  The curve labelled \emph{Tidal+Rotation+Resonant
  torque} (dotted) includes all three mechanisms: tidal truncation,
rotation (both the centrifugal and Coriolis effects), and the resonant
torque.

We also plot the mass loss from the circular-orbit N-body simulation
in Fig. \ref{fig:massloss.iso} (solid line).  The idealised
simulation including all the processes matches the actual N-body
simulation remarkably well.  However, the choice of $\alpha=0.75$,
although quite reasonable, was rather arbitrary and we made this
choice to match the circular orbit simulation.  But Fig.
\ref{fig:massloss.iso} clearly shows that the resonant torque causes
more than 70\% of the satellite mass loss.  The remaining 30\% of the
mass loss results from rotational effects. Tidal truncation does not
cause much satellite mass loss.

\begin{figure}
  \centerline{\includegraphics[width=1.0\columnwidth,angle=0]
    {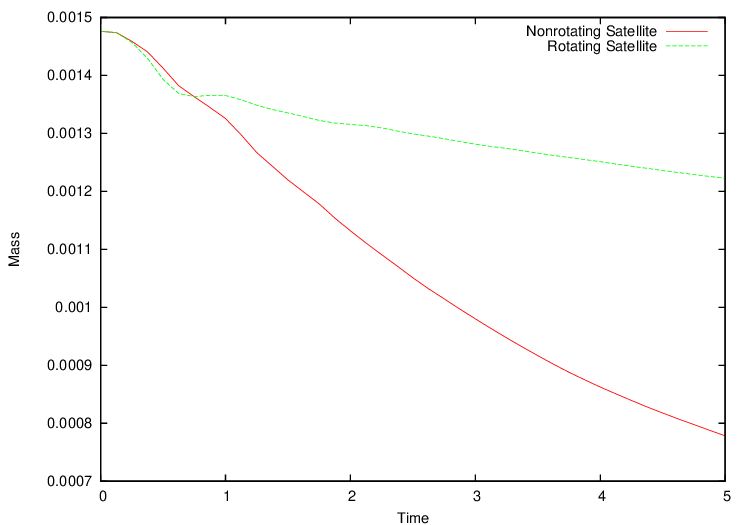}}
  \caption{The mass loss history of circular orbit simulations for
    both a non-rotating and a rotating satellite. Since the rotating
    satellite simulation suppresses the resonant torque, its mass loss
    is significantly reduced.}
  \label{fig:massloss.rot}
\end{figure}

A satellite whose frame corotates with its revolution on a circular
orbit should not experience a resonant torque.  So as a further test of
the resonant torque mechanism, we compare the mass loss from a
non-rotating satellite simulation to a rotating satellite simulation.
The non-rotating satellite simulation is the same as in \S
\ref{sec:res_torque}. For the rotating satellite simulation, the
satellite particles are rotated every time step around an axis
perpendicular to the orbital plane, which passes through the
satellite's centre. They are rotated with the same frequency as the
circular orbit frequency of the satellite in the host halo.  This
fixes the direction of the host halo centre in the satellite frame.
Owing to this rotation, the halo does not feel a time-dependent
external potential, and hence feels no resonant torque. The two
simulations have the same initial conditions. However, owing to the
artificial rotation, the rotating satellite feels an extra Coriolis
force.  During the course of rotating satellite simulation, we
subtract this extra force to make the comparison more meaningful.

Fig. \ref{fig:massloss.rot} shows that the rotating satellite loses
only $\approx40$\% of the mass lost by the non-rotating satellite by
$T=5.0$. Moreover, the mass loss histories of both satellites are
almost identical at early times ($T \leq 0.7$). This mass loss results
from the adjustment of the satellite's initial conditions to the host
halo potential. If we ignore satellite mass loss during this phase,
the non-rotating satellite loses about five times more mass than the
rotating satellite, again confirming that resonant torques are the
dominant mechanism responsible for circular orbit satellite mass loss.

The mass loss rates in the \emph{Tidal+Rotation} simulation in Fig.
\ref{fig:massloss.iso} and the \emph{Rotating} satellite in Fig.
\ref{fig:massloss.rot} are almost identical if we ignore the early
relaxation phase in the rotating satellite simulation. Both
simulations attempt to reproduce circular orbit satellite mass loss
without resonant torque effects, using totally different
approaches. This similarity confirms that our treatment for
non-resonant mass loss captures the essential dynamics and implies
that that our estimates of satellite mass loss in these experiments is
reasonable.  In conclusion, Figs. \ref{fig:massloss.iso} and
\ref{fig:massloss.rot} confirm that resonant torque effects dominate
circular orbit satellite mass loss.

\section{Satellite disruption on an eccentric orbit}
\label{sec:ecc}

\begin{figure}
  \centerline{\includegraphics[width=1.0\columnwidth,angle=0]
    {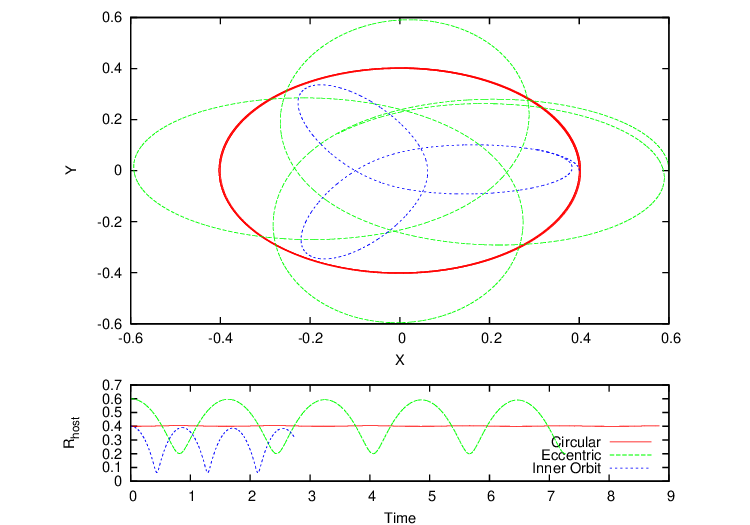}}
  \caption{The circular, eccentric, and inner orbits.  \emph{Top
      panel:} the three trajectories in the orbital plane.
    \emph{Bottom panel:} the time evolution of the distance between
    the satellite and the host halo centre.}
  \label{fig:orbit.X3}
\end{figure}

Several physical processes, including resonant torques, gravitational
shocks with resonant shock effects, and continuous tidal truncation,
simultaneously affect a satellite on an eccentric orbit.  To study
this complicated interplay and to understand the characteristics of the
individual physical processes, we compare the evolution of three
different simulations: a circular orbit simulation, an eccentric orbit
simulation, and an inner orbit simulation (see Fig.
\ref{fig:orbit.X3}).  The circular orbit simulation is that described in
\S\ref{sec:circ}.  The eccentric orbit simulation is an $e=0.5$ orbit
with an apocentre of $r_{apo}=0.6$ and a pericentre of $r_{peri}=0.2$
\footnote{We define eccentricity as $e \equiv
  (r_{apo}-r_{peri})/(r_{apo}+r_{peri})$}.  The circular orbit and the
eccentric orbit have the same orbital energy.  The inner orbit
simulation is an $e=0.72$ orbit with an apocentre of $r_{apo}=0.4$ and
a pericentre of $r_{peri}=0.064$.  The normalised angular momentum
($\kappa$) of the inner orbit is $0.55$, which is near the median
$\kappa$ of subhaloes in a sample taken from recent cosmological
simulations \citep{Ghigna98,Zentner05}.

\begin{figure}
  \centerline{\includegraphics[width=1.0\columnwidth,angle=0]
    {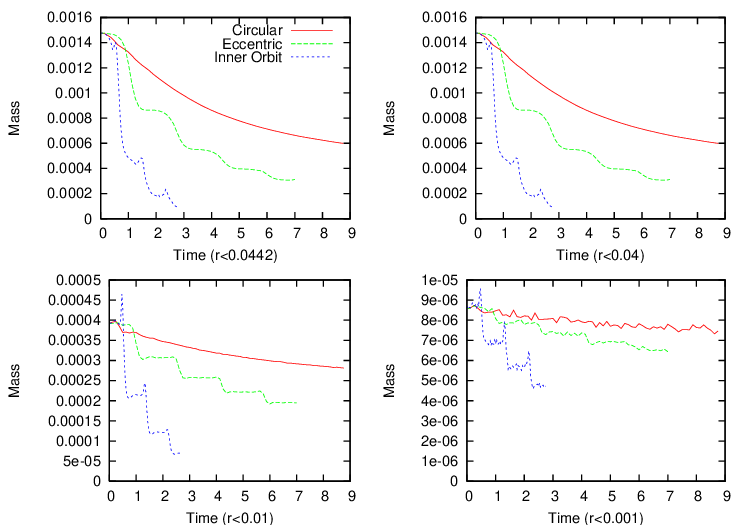}}
  \caption{The mass evolution for the circular, eccentric, and inner
    orbit simulations. We plot the mass within radii of 0.045, 0.04,
    0.01, and 0.001 versus time in the top left, top right, bottom
    left, and bottom right panels, respectively.  The initial
    satellite radius is 0.045.}
  \label{fig:massloss.X3}
\end{figure}

Fig. \ref{fig:massloss.X3} shows the evolution of the enclosed mass
within 90\%, 22\%, 2.2\%, and 0.22\% of the original satellite radius.
The two eccentric orbit satellites lose their mass at pericentre
episodically, while that for the circular orbit appears continuous.
The eccentric-orbit satellites lose significantly more mass than the
circular orbit satellite.

\begin{figure}
  \centerline{\includegraphics[width=1.0\columnwidth,angle=0]
    {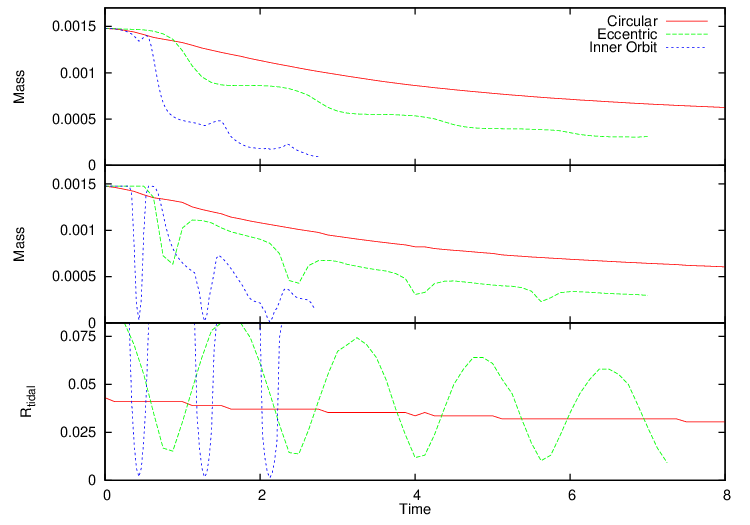}}
  \caption{Evolution for the circular, eccentric and inner orbit
    simulations.  \emph{Top panel}: the mass enclosed within the
    initial satellite radius.  \emph{Middle panel}: the mass enclosed
    within the instantaneous tidal radius.  \emph{Bottom panel}: the
    evolution of the tidal radius.  }
  \label{fig:massloss.Rt}
\end{figure}

The computation of a satellite's gravitationally bound mass is
difficult in practise and we use the mass enclosed within the
truncation radius as a proxy.  Fig. \ref{fig:massloss.Rt} shows the
mass loss history for the three simulations using two different
definitions of the bound satellite mass.  The first definition is the
total mass inside the initial satellite radius.  The second definition
is total mass inside the instantaneous tidal radius.  We find that a
satellite's mass rapidly decreases and then increases again around the
time of pericentre passage, if we use the tidal radius definition for
the bound mass.  This mass evolution results from the rapid variation
in the instantaneous tidal radius (Fig.  \ref{fig:massloss.Rt},
bottom).  However, if this radius varies faster than the orbital time
of an escaping particle, the tidal radius will not represent the
dynamics of escape.  Therefore, we generally use the initial satellite
radius definition for our working definition of the bound satellite
mass.

\subsection{Heating by the gravitational shock at pericentre}
\label{sec:peri}

According to Fig. \ref{fig:massloss.X3}, satellites on eccentric
orbits lose mass mostly at pericentre.  This suggests that satellite
heating by the gravitational shock at pericentre is the dominant
mechanism for satellite mass loss and disruption.  We can discriminate
two distinct effects of this gravitational shock: evolution of the
satellite's structure and the persisting tidal truncation.

\begin{figure*}
  \centerline{
    \includegraphics[width=0.825\textwidth,angle=0]{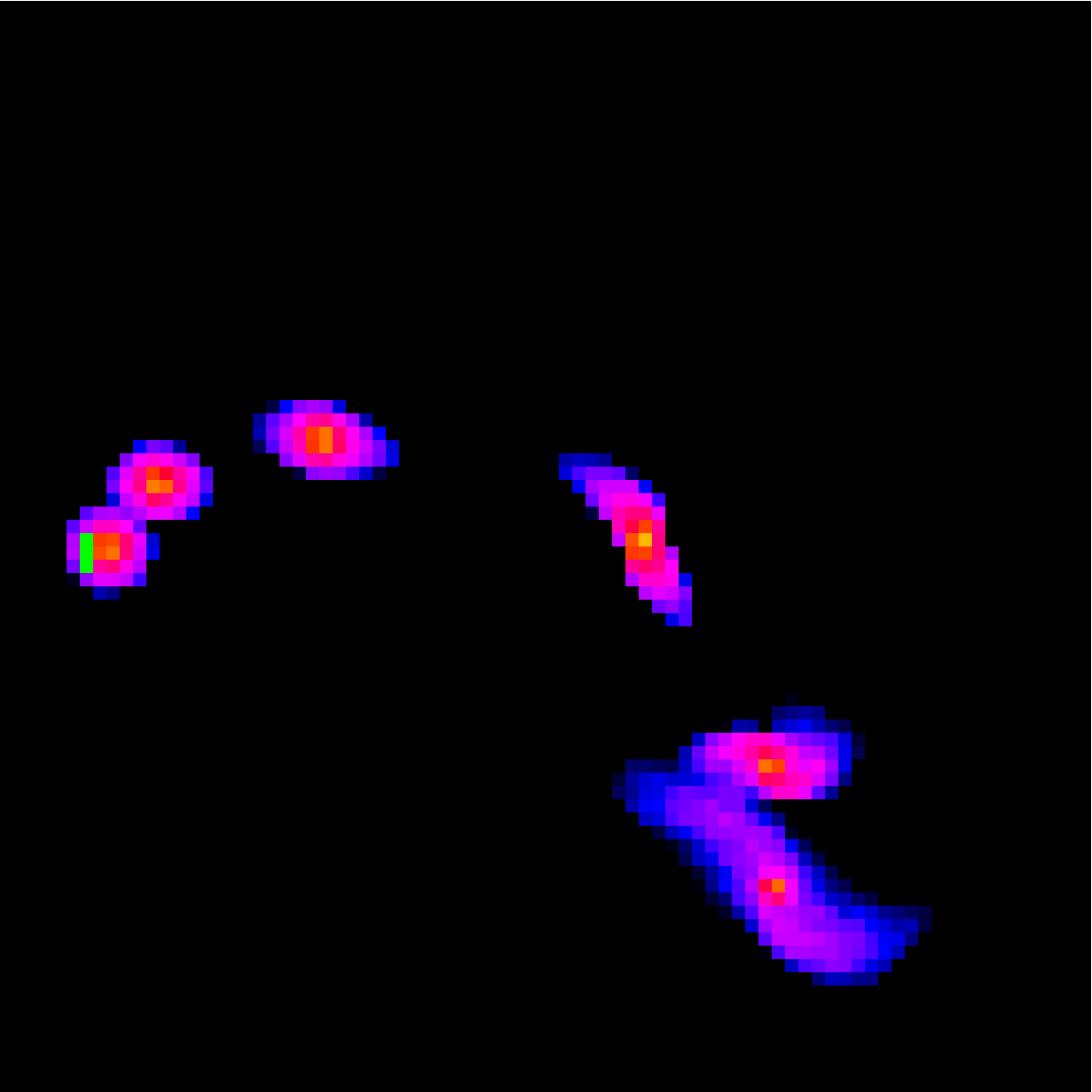}
  } \centerline{
    \includegraphics[width=0.41\textwidth,angle=0]{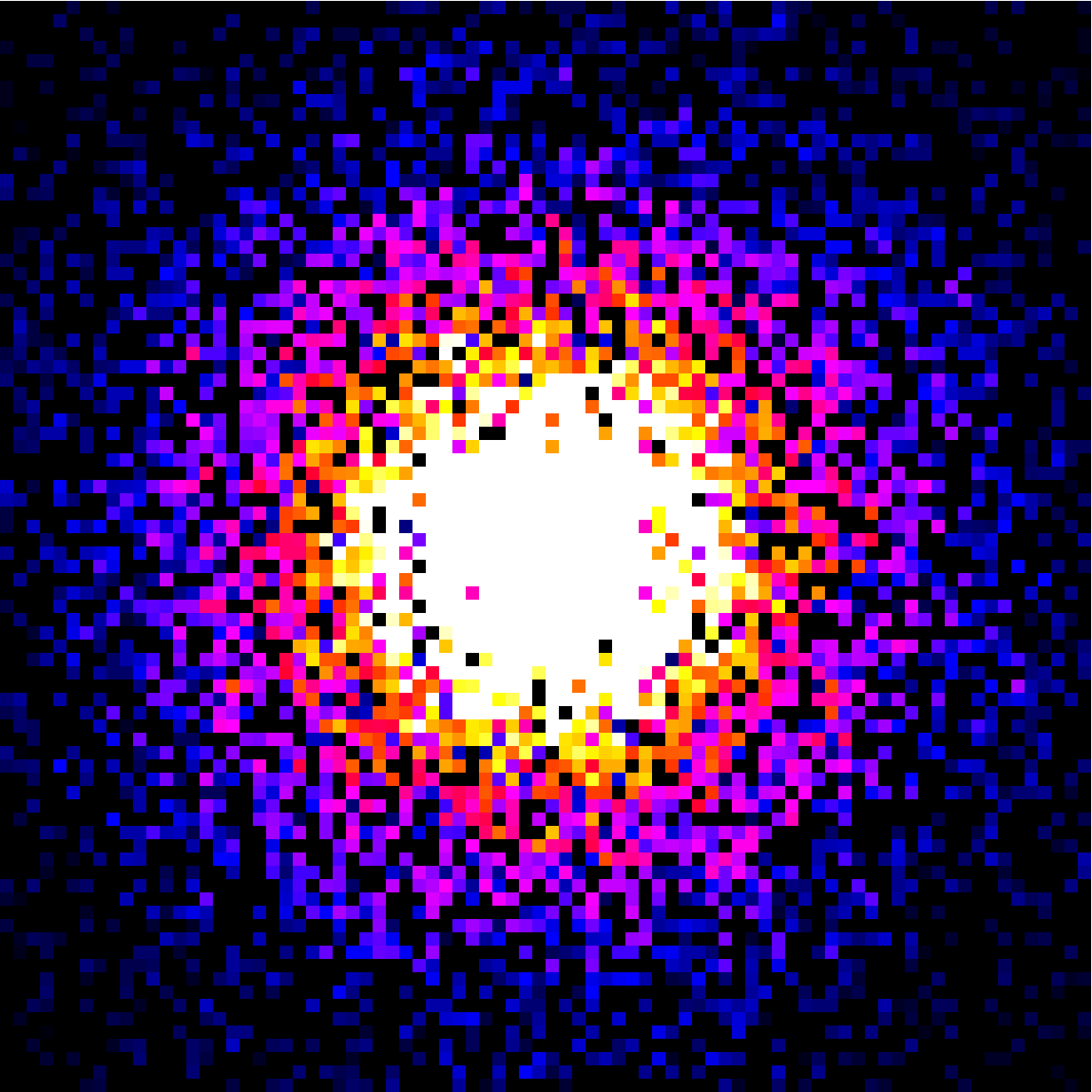}
    \includegraphics[width=0.41\textwidth,angle=0]{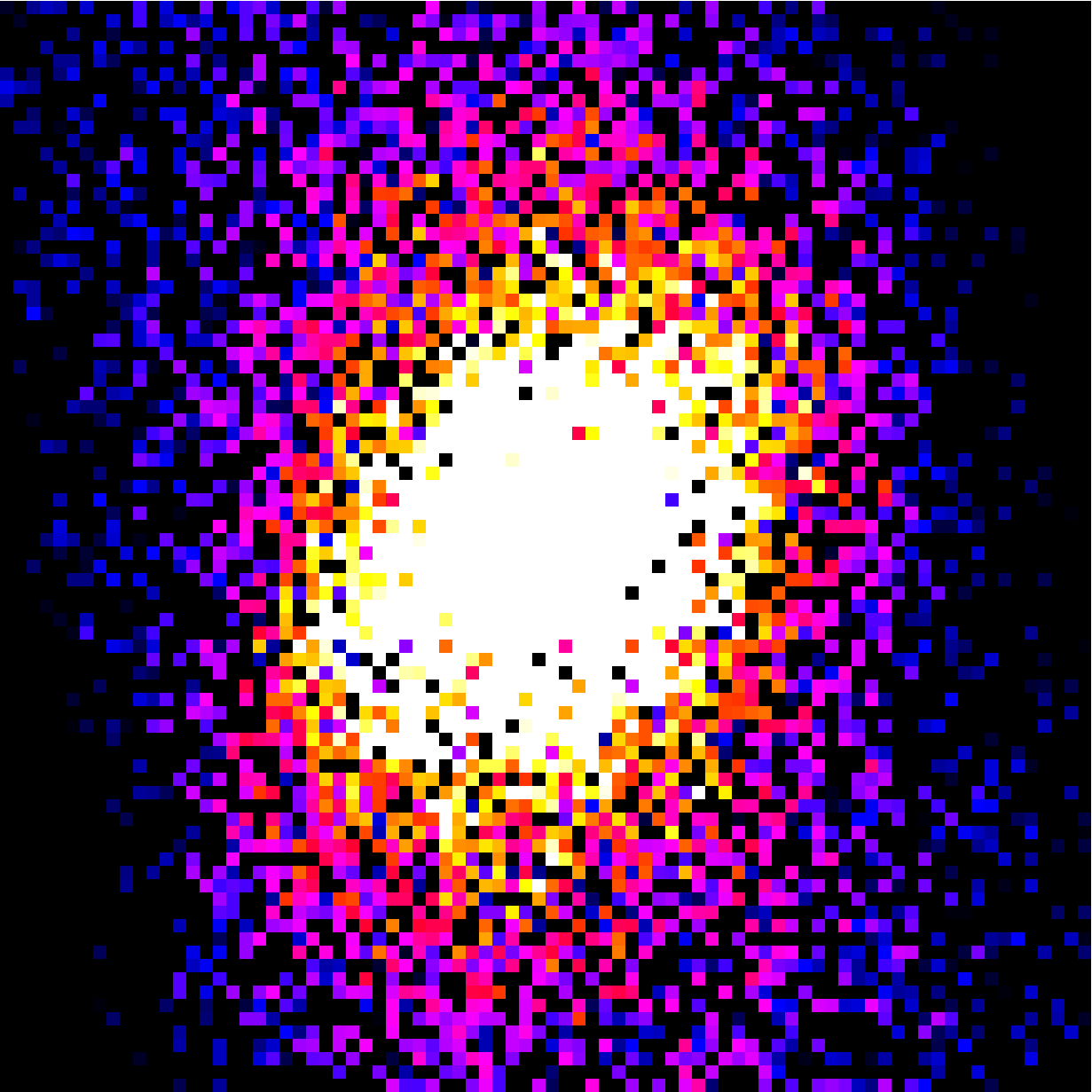}
  }
  \caption
   {Snapshots of the inner orbit simulation for one radial period
    beginning at apocentre.  {\it Top panel:} Overlapping snapshots of
    the satellite, shown at $T=$0.0, 0.16, 0.31, 0.47, 0.63,
    and 0.78.  The origin of the inner axes (green) is the centre of the host
    halo.  The dark matter density scales logarithmically with
    colour.  {\it Bottom panels:} The
    satellites at $T=0.0$ and $T=0.31$ zoomed in to the size of the
    small boxes (green) in the top panel.  To emphasise the dark matter density
    differences, the colour scale is now linear in density.  The bottom
    snapshots show an increase in the size of the high density regions at
    $T=0.31$ at pericentre compared to the satellite at $T=0.0$ at apocentre.}
  \label{fig:Frames.XK2}
\end{figure*}

The top panel in Fig. \ref{fig:Frames.XK2} shows overlapping
snapshots of the inner orbit simulation over one orbital period
beginning from apocentre.  The colour scale is logarithmic in
satellite density. The lower-right panel shows the satellite evolution
during pericentre passage.  The satellite is compressed
perpendicular to its trajectory and stretched out along its trajectory
by the external potential.  This deformation causes the density to
increase, as observed in Fig. \ref{fig:massloss.X3} where the
enclosed mass within 90\% (the top right panel), 22\% (the bottom left
panel), and 2.2\% (the bottom right panel) of the original satellite
radius increases at pericentre.  The two bottom panels in Fig.
\ref{fig:Frames.XK2} show blown-ups of the two boxed areas in the top
panel.  To highlight density differences, the colour now scales
linearly with satellite density from $\rho =1$ to $\rho=100$, and
white represents $\rho > 100$.  The area of the white region in the
right panel is larger than that of the left panel because of the
density enhancement at pericentre.

The energy input from the gravitational shock reduces the satellite's
binding energy, causing it to expand, and leads to tidal stripping.
Work is done on the satellite potential to strip particles, this extra
energy drives expansion and enhances the mass loss.  In addition to
enhancing the mass loss, the tidal truncation causes the satellite
haloes to lose equilibrium.  Fig. \ref{fig:massloss.Rt} shows
that when the satellite moves deep inside the host halo, the tidal
radius decreases.  A large fraction of the satellite material outside
of the tidal radius is not stripped and reenters the satellite when it again
moves outward in the host halo and the tidal radius again increases.
This rapidly changing tidal radius keeps the satellite haloes out of
equilibrium.

\begin{figure}
  \centerline{\includegraphics[width=1.0\columnwidth,angle=0]
    {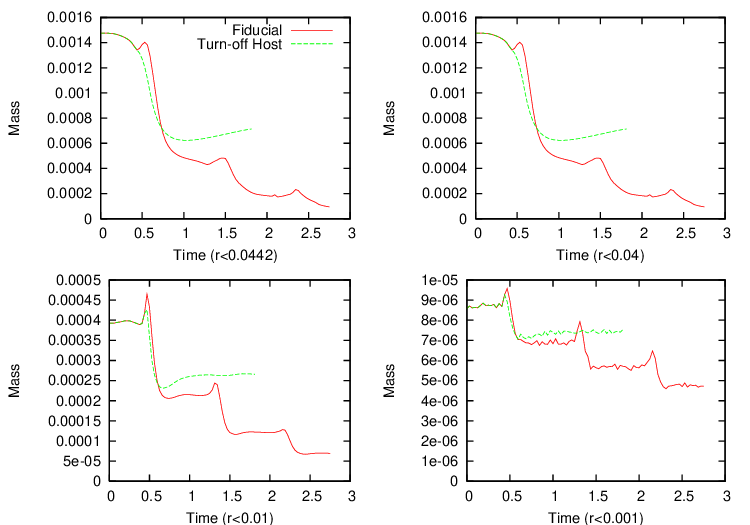}}
  \caption{The figures show the enclosed mass evolution of two
    simulations: the fiducial simulation, which is the same as the
    inner orbit simulation, and the `turn-off' host simulation, where
    the host halo potential is turned off at pericentre.}
  \label{fig:massloss.XK_TO}
\end{figure}

To help understand the effect of the host halo potential, we compare
two inner orbit simulations. The first simulation (fiducial) is the
same as the original inner orbit simulation.  The second simulation
(turn-off host) is the same inner orbit simulation but we artificially
``turn off'' the host halo potential at pericentre ($T \approx 0.45$).
We present the mass loss histories of these two simulations in Fig.
\ref{fig:massloss.XK_TO}. After turning off the host halo potential,
the mass loss history of the two satellites diverges.  The turn-off
host simulation shows $\sim$ 15\% less mass loss than fiducial
simulation.  In addition, the satellite mass loss history of the
turn-off host simulation shows an increase in mass after the
significant mass loss caused by the gravitational shock, while the
mass loss history of the fiducial simulation shows a continuous mass
loss after the shock.  The increase in mass results from lost
particles returning to the satellite.  Conversely, the persisting
tidal truncation by the host halo potential in the fiducial simulation
enhances satellite mass loss, causing
the satellite halo to lose equilibrium and makes the evolution
nonlinear.

In summary, both the internal structural evolution and the subsequent
tidal truncation driven by time-dependent heating leads to significant
evolution.  These nonlinear physical processes are essential
ingredients for an accurate prediction of satellite disruption and,
therefore, simple approximations such as impulsive heating are
insufficient \citep{Spitzer87}.

\subsection{Resonant heating on an eccentric orbit satellite}
\label{sec:res_ecc}

We have divided the resonant dynamics of satellite evolution into a
shock and a torque, but this distinction is arbitrary in principle.
For example, both resonant effects simultaneously drive the evolution
of a satellite on an eccentric orbit.  A consistent perturbation
theory calculation for an eccentric orbit perturber incorporates the
full time dependence, coupling both the radial and the azimuthal
orbital frequencies to a satellite's phase space.  In addition,
orbital decay adds the complication of a continuous spectrum of
perturbation frequencies.  Hence, a perturbation theory calculation
for an eccentric orbit is a difficult and expensive.  To get around
this difficulty, we estimate the response by considering the most
important frequencies one at a time.  The strongest resonant effects
are associated with the pericentric passage.  We can, therefore,
approximate the coupling with that of a circular satellite orbit at
the pericentre radius and proceed as in \S\ref{sec:res_torque}.

\begin{figure}
  \centering
  \includegraphics[width=1.0\columnwidth,angle=0]{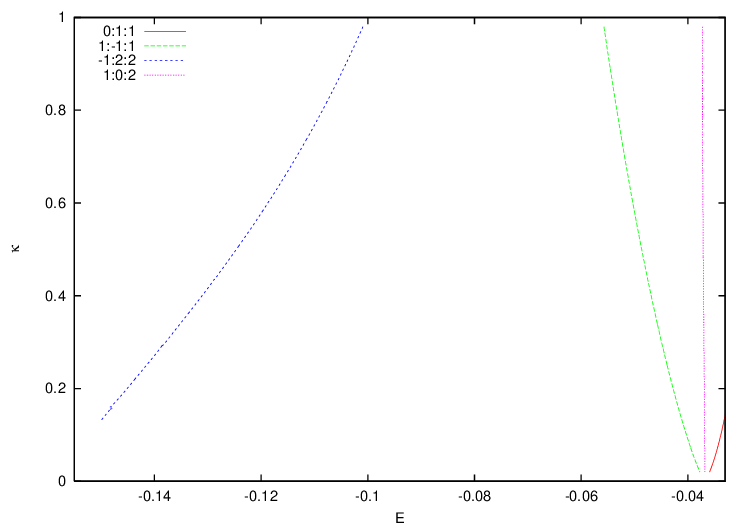}
  \caption{As in Fig. \protect{\ref{fig:FindRes_XT}} but assuming a satellite
    frequency of $\Omega_{sat}=6.56$,  which corresponds to the
    frequency of a circular orbit at $r_{peri}$ for the eccentric
    orbit simulation.  Only the $l=1$ and $l=2$ resonances are shown.
    Unlike in Fig. \ref{fig:FindRes_XT}, the 1:-1:1, 0:1:1, and
    1:0:2 resonances are located inside the satellite.}
  \label{fig:FindRes_XT_Ecc}
\end{figure}

Fig. \ref{fig:FindRes_XT_Ecc} shows the locations of the resonances
in satellite phase space for this equivalent circular orbit problem.
We examined resonances with $l_1\in[-10, 10]$ and $l\in[1,4]$.  In
addition to the -1:2:2 resonance, the 0:1:1, 1:-1:1, and 1:0:2
resonances occur within the satellite's phase space.  Comparing this
to the results from the larger radius circular orbit calculation
(\S\ref{sec:res_torque}), the coupling of the larger instantaneous
angular frequency at pericentre to higher binding energy orbits that
have larger orbital frequencies, moves the resonances inward in
radius.  Hence, some resonances that were previously outside the
satellite in \S\ref{sec:res_torque} are now within the satellite.  In
addition, the stronger tidal field at smaller orbital radii in the
host increases the strength of the coupling.  Using numerical
perturbation theory, the strength of the 1:-1:1 and the 1:0:2
resonances are now comparable to the strength of the -1:2:2 resonance,
which is much stronger than the -1:2:2 resonance in the larger radius
circular orbit.

In summary, resonant heating is enhanced for an eccentric orbit both
because of the larger number of resonances within the satellite and
the stronger tidal force felt by the satellite at pericentre.
However, our approximate perturbation theory
calculation does not include the full time dependence of the eccentric
orbit, a calculation that is extremely difficult and is beyond the
scope of this paper.

\section{Satellite stripping and its effects}
\label{sec:stripping}

\subsection{The stripping process}
\label{sec:strip_proc}

Owing to satellite heating, particles in a satellite gain energy and
angular momentum.  This reduces the satellite's binding energy and
enhances the satellite's mass loss.  There are several mechanisms that
heat a satellite and each one is effective over different ranges
of binding energy.  However, the re-equilibration of the satellite
tends to globally redistribute the work throughout the satellite
profile, washing out the nature of its origin.  The observational
signatures of the subsequent mass loss, then, tend to be universal.

\begin{figure}
  \centering
  \includegraphics[width=1.0\columnwidth,angle=0] {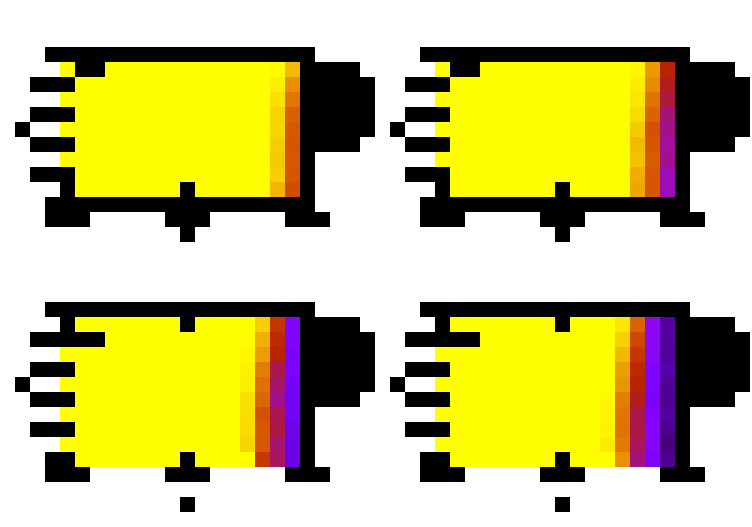}
  \caption{The fraction of particles remaining in phase space at
    different times.  The colour code is linear from all particles
    stripped (0) to no particles stripped (1).  Each panel shows the
    fraction of particles remaining at the labelled time compared to
    the initial satellite.  Satellite stripping is an outside-in
    process in energy space.}
  \label{fig:2DPSP.XC1}
\end{figure}

\begin{figure}
  \centering
  \includegraphics[width=1.0\columnwidth,angle=0]{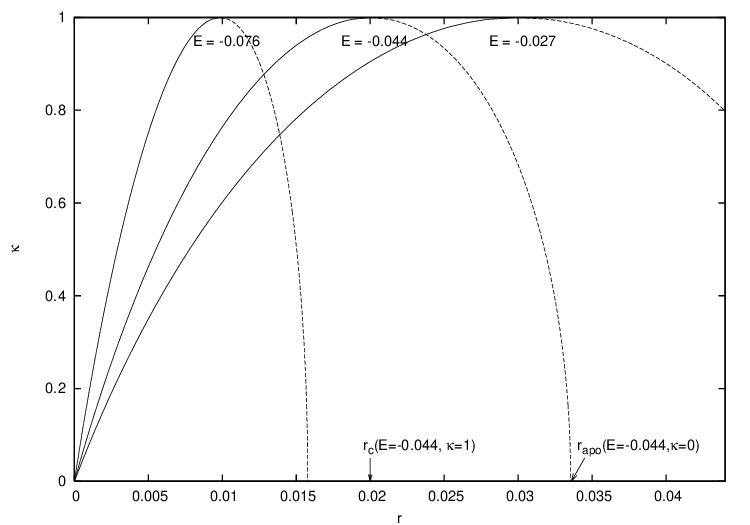}
  \caption{The apocentre and pericentre of orbits in the initial
    satellite model.  Dashed lines represent the apocentre and solid
    lines the pericentre for orbits with energies of $E=$-0.076,
    -0.044, and -0.027.  The radii for circular orbits with these
    energies are 0.01, 0.02, and 0.03, respectively.  The x-axis is
    the distance from the halo centre and the y-axis is the normalised
    angular momentum, $\kappa$.  The area between the pericentre and
    apocentre lines is the range in radii for orbits with each $E$ and
    $\kappa$. The two arrows show the circular orbit radius and the
    apocentre radius for a radial orbit ($\kappa=0$) with $E=$-0.044.}
  \label{fig:apo_peri}
\end{figure}

Fig. \ref{fig:2DPSP.XC1} plots the fraction of bound particles
remaining in different parts of phase space at different times for a
satellite on a circular orbit.  Low binding energy particles are
stripped at earlier times and high binding energy particles are
stripped later.  In other words, the satellite stripping process is an
outside-in process in energy space.  We can understand this behaviour
as follows.  Weakly bound satellite orbits are affected and stripped
by the tidal force beyond a characteristic radius.  The apocentres of
orbits with a given energy are within a factor of two of the radius of a
circular orbit with the same energy, even for a zero-angular momentum
orbit. For example, Fig. \ref{fig:apo_peri} marks the radius of a
circular orbit ($\kappa=1$) and the apocentre of a radial orbit
($\kappa=0$) for an energy of $E=-0.044$; the apocentre of the radial
orbit is only 50\% larger than the radius of the circular orbit.
Therefore, energy and not the relative angular momentum $\kappa$
determines the stripping boundary. Although very low angular momentum
orbits can be stripped at lower energies, Fig. \ref{fig:2DPSP.XC1}
shows that any trend towards a larger escape fraction for smaller
$\kappa$ at fixed energy is very weak.

\begin{figure}
  \includegraphics[width=1.0\columnwidth,angle=0] {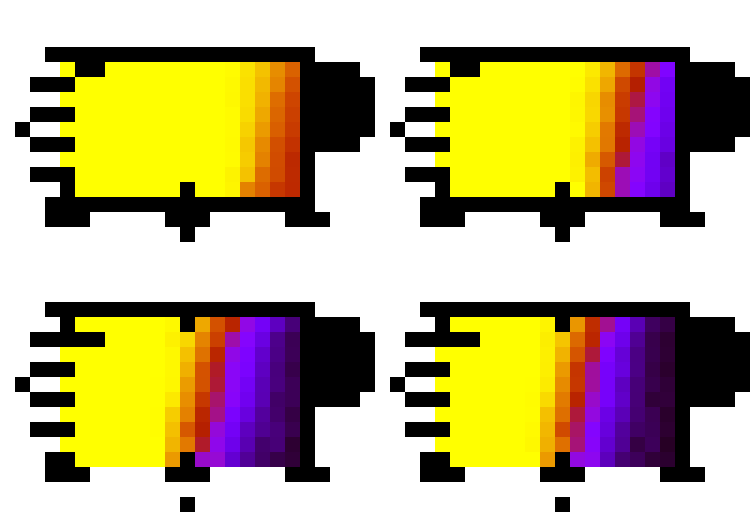}
  \caption{The same as Fig. \ref{fig:2DPSP.XC1} but for the
    eccentric orbit simulation.}
  \label{fig:2DPSP.XE1}
\end{figure}

We plot the fraction of particles remaining in different parts of
phase space for the eccentric orbit simulation in Fig.
\ref{fig:2DPSP.XE1}.  The stripping process remains an outside-in
process in energy space but now the trend toward a larger escape
fraction at smaller relative angular momentum is stronger.  This
angular momentum dependence results from the time dependence of the
tidal radius as shown in Fig. \ref{fig:massloss.Rt}.  Near a
satellite's pericentre, satellite particles with low angular momentum
are temporarily outside the instantaneous tidal radius and are prone
to escape.

The outside-in stripping in energy space has interesting consequences
for satellite evolution.  Since the stars and cold gas in a satellite
galaxy are centrally concentrated in their own dark matter halo, it
protects them from stripping.  The degree of \emph{protection} for any
component depends on its binding energy.  For example, consider a disk
and elliptical galaxy with the same linear extent\footnote{In fact
  elliptical galaxies are generally more concentrated than disk
  galaxies.} and the same orbit in the host halo. The disk galaxy
contains material mostly on circular orbits, which have the largest
binding energy at fixed radius, while the elliptical galaxy is made
from a wide range of stellar orbits.  We therefore predict that stars
in the elliptical galaxy will be stripped sooner than stars in the
disk galaxy.  This difference should be incorporated into models of
satellite galaxy stripping and when one estimates the subsequent stellar
halo distribution.

\subsection{The LMC stellar tail}
\label{sec:LMC}

The LMC is the largest Milky Way satellite.  It has been the subject
of numerous observational studies and characteristics of its structure
and kinematics have been well established.  However, the origin of the
Magellanic Stream, a thin neutral hydrogen tail stretching over
$100^{o}$ along a Galactic great circle, remains unexplained.
Observations suggest that the Magellanic Stream may be a relic of a
past interaction with the Milky Way \citep{MSM77,Putman98}.  If this
is the case, tidally stripped stellar ejecta should be observed as
predicted by tidal interaction theory.  However, none of the many
searches for stellar ejecta has successfully detected a population of
stars connected with the Magellanic Stream.  Several scenarios have
been suggested to resolve this conflict. First, the LMC gas may be
removed by ram-pressure forces caused by an interaction with the outer
disk \citep{MD94} or with hot gas in the Galactic halo
\citep{Mastropietro05} instead of a tidal interaction.  Second, the
Stream may be a remnant from the tidal disruption of the Small
Magellanic Cloud (SMC) owing to its gravitational interaction with the
LMC and the Milky Way \citep{MF80,Connors06}.  Based on the results
from the previous sections, we suggest that the LMC stars are
protected by its dark matter halo while the Stream is the remnant of a
low binding energy, extended neutral gas disk, analogous to the Milky
Way's extended H{\sc I} disk. Although smaller satellites and globular
clusters have symmetric tidal tails, which has been used as an
argument against a tidal origin for the Magellanic Stream \citep{MD94},
larger satellites have a asymmetric tidal tails owing to the asymmetry
in the tidal force \citep{Choi.etal:07}. After the gas is stripped it would
then hydrodynamically interact with the hot halo gas, changing the orbit of the
gas in the stream \citep{Mastropietro05}.

\begin{figure}
  \centerline{
    \includegraphics[width=1.0\columnwidth,angle=0] {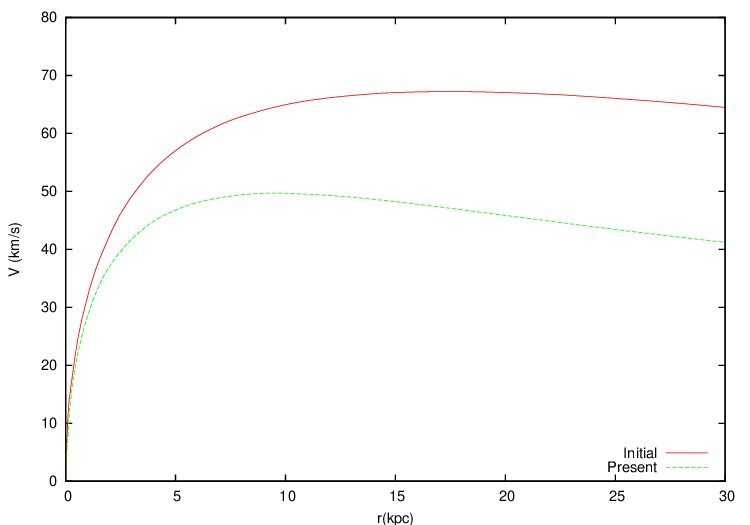}
  }
  \caption {The circular velocity profile of the LMC dark matter halo at
    the initial and present time.  }
  \label{fig:lmc.profile}
\end{figure}

We implement an idealised simulation of LMC evolution in the Milky
Way dark matter halo to test this hypothesis.  The static Milky Way
dark matter halo potential is based on the A1 model of \citet{KZS02};
a $c=12$ NFW halo with $R_{vir}=258$ kpc, $M_{vir}=1.0 \times
10^{12} M_{\sun}$, and $V_{max} = 163\,\hbox{km/s}$.  The live LMC
halo model, which is also based on a $c=12$ NFW halo model, is truncated at
its $x_{e}$ radius assuming that the LMC is located at $0.6R_{vir}$.
To make a stable halo model we performed an Eddington inversion to the
truncated LMC halo model, as presented in \S\ref{sec:method}, and we
realise a phase space of $10^{6}$ equal-mass particles. The initial
LMC halo circular velocity, $V_{max, LMC}=0.42V_{max, host}$, is
chosen to match the observed LMC circular velocity of $V_{max,
  LMC}=0.3V_{max, host}$ \citep{Kim98,vMAHS02} at the present day 
(see Fig. \ref{fig:lmc.profile}).

Since the LMC's evolutionary history is not
well-defined, we assume a simple LMC orbit: an $e=0.5$ orbit with a
pericentre at $0.2R_{vir}$ and an apocentre at $0.6R_{vir}$.  The
current LMC is assumed to be located just after its pericentre
\citep{GSF94,GN96}.  The simulation starts at $0.6R_{vir}$ and 
runs a little more than two radial periods to represent the present-day 
LMC. It is worth noting that a consensus on
the LMC orbital parameters has not yet been reached. Recent proper 
motion measurements using the Hubble Space Telescope
\citep{Kallivayalil.etal:2006} imply a very large galactocentric velocity. 
This large LMC velocity combined with an improved Milky Way mass model
suggests either that the LMC is on its first passage about the Milky Way or
that its orbital period and apocentre must be a factor of 2 larger than
previously estimated \citep{Besla.etal:2007}.  In this paper, we take the
classical LMC orbital parameters, which are based on the 
prescription induced by \citet{MF80}.  If it turns out that the classical 
point of view is false, our quantitative estimation should be revisited.
However, our qualitative interpretation about the pattern of the LMC 
disruption is not sensitive to our particular choice of LMC orbital 
parameters. 

\begin{figure}
  \centerline{\includegraphics[width=0.85\columnwidth,angle=0]{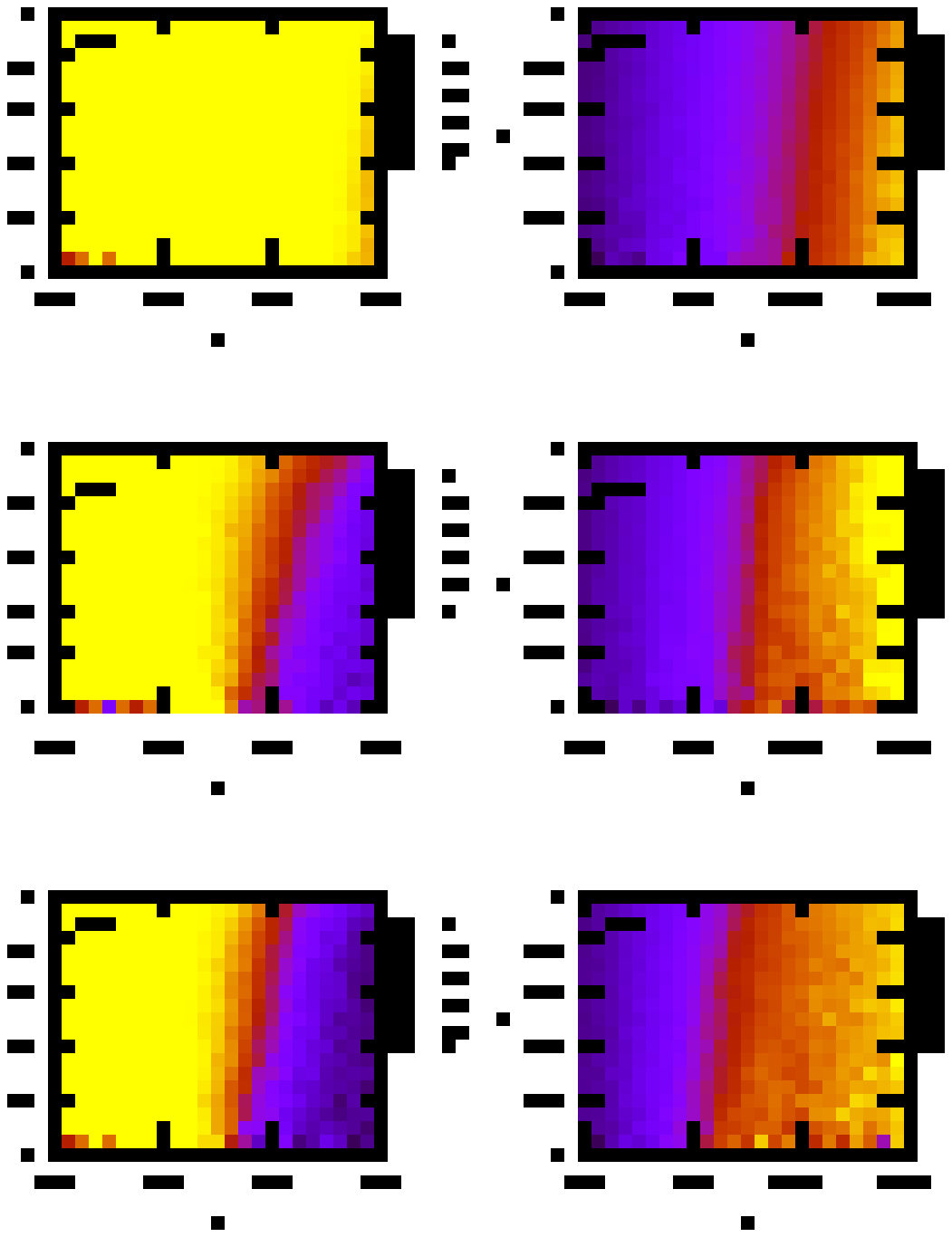}}
  \caption{The distribution of particles remaining in the LMC halo at
    different times.  {\it Left panels:} The fraction of particles
    remaining in the LMC halo based on their initial location in phase
    space (see Fig.  \protect{\ref{fig:2DPSP.XC1}}). {\it Right
      panels:} The mean radius of the particles remaining as a
    function of phase space coordinates.}
  \label{fig:2DPSP.LMC}
\end{figure}

Fig. \ref{fig:2DPSP.LMC} shows the fraction of particles remaining
in the LMC in different parts of phase space (left panels) and their
average distance from the LMC centre (right panels) at different
times. Once again, the stripping corresponds to an outside-in process
in energy space.  Fig. \ref{fig:2DPSP.LMC} shows that particles with
$E < -0.5$ have not yet been stripped.  We assume that the LMC stars are
mostly on circular orbits ($\kappa > 0.9$), because the visible LMC is
a rotationally supported system. The current mean radius of the
particles for $E=-0.5$ and $\kappa = 0.9$ is about 15 kpc, and hence LMC
stars inside 15 kpc are expected to be protected by the dark matter
halo, which is larger than the observed disk radius of 10 kpc \citep{Kunkel97}. 
This suggests that the detectable
signature of the LMC stellar ejecta could be very small.

\subsection{Density profiles and rotation curves}
\label{sec:profile}

\begin{figure*}
  \subfigure[The circular orbit]{
    \includegraphics[width=0.45\textwidth,angle=0]{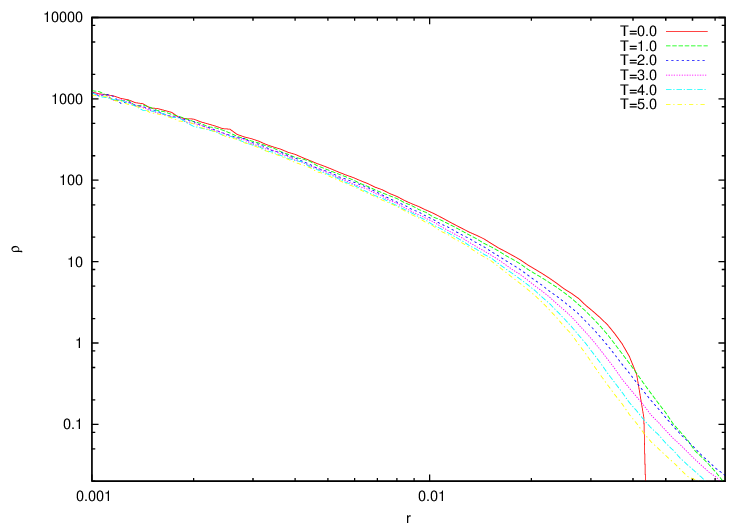}
    \includegraphics[width=0.45\textwidth,angle=0]{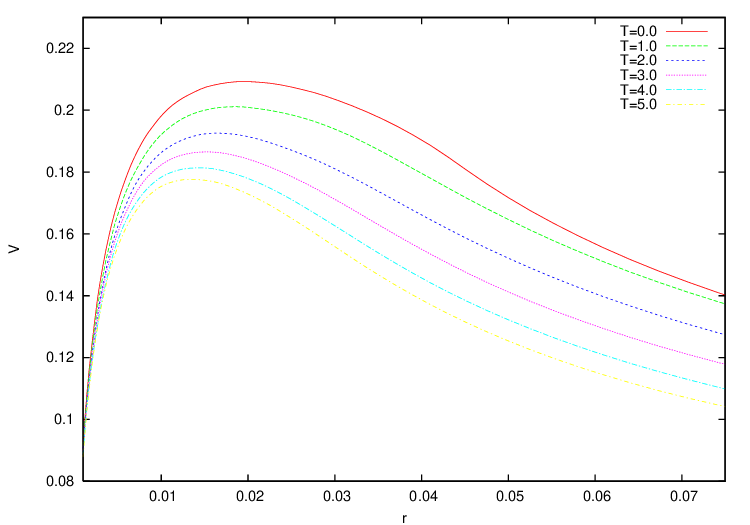}
  } 
  \subfigure[The eccentric orbit]{
    \includegraphics[width=0.45\textwidth,angle=0]{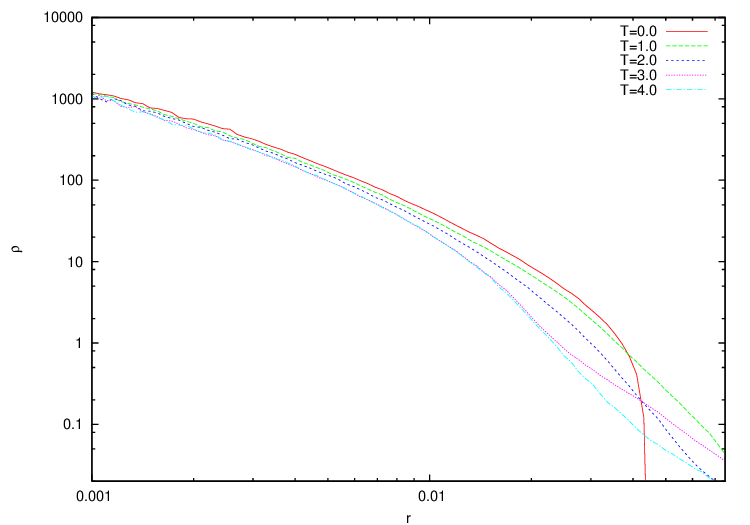}
    \includegraphics[width=0.45\textwidth,angle=0]{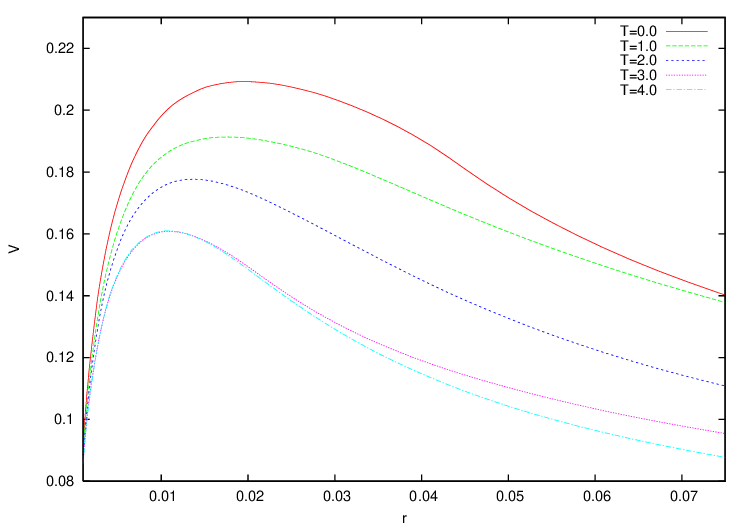}
  } 
  \subfigure[The inner orbit]{
    \includegraphics[width=0.45\textwidth,angle=0]{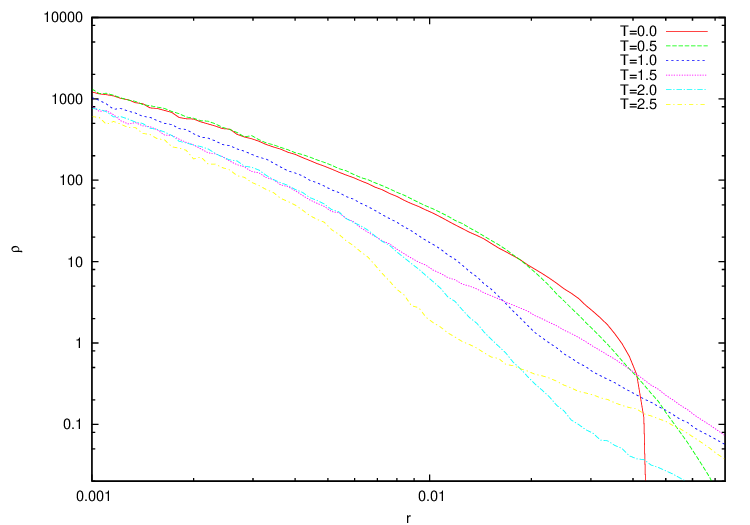}
    \includegraphics[width=0.45\textwidth,angle=0]{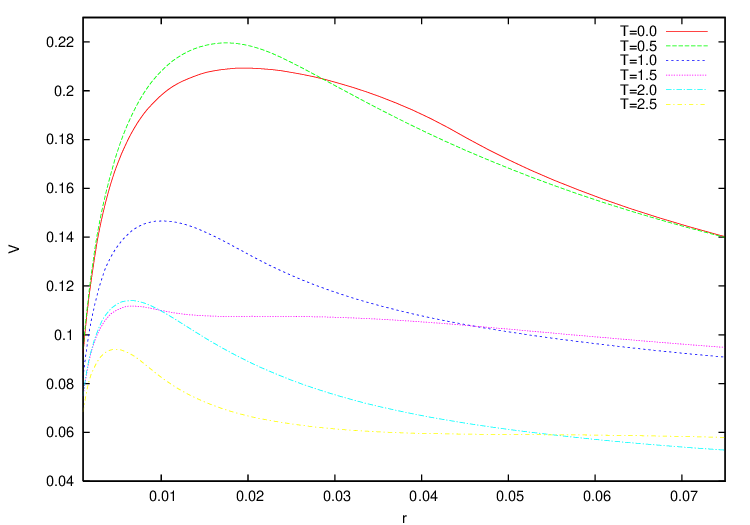}
  }
  \caption{The density profiles (left panels) and 
    rotation curves (right panels) of the
    satellite dark matter haloes for the three satellite orbit
    simulations at various times.}
  \label{fig:profile.evolution}
\end{figure*}

The left panels in Fig. \ref{fig:profile.evolution} show the evolution
of the density profiles for the three different orbit simulations.  As
expected given the mass loss rates shown in Fig.
\ref{fig:massloss.X3}, the inner orbit satellite's density profile
evolves the most and the circular orbit's density profile evolves the
least.  However, even though the mass loss rates are different, the
overall trends in the density profile evolution for the three
satellites are similar.  A decrease in density occurs at all radii,
which is particularly evident in the inner orbit simulation, where
even at $r \le 0.001$ the density decreases uniformly with time.
Although the density decreases, the steepness of the density profile
does not decrease. Hence, the inner cusp is preserved during tidal
mass loss, i.e. the evolution of satellites owing to interactions with
their host halo will not remove NFW central cusps to make them agree
more with observed density profiles
\citep{Diemand.etal:07,Springel.etal:08}.

One can see the effect on the circular velocity profiles for the
satellites on the three different orbits in the right panels of Fig.
\ref{fig:profile.evolution}.  Once again, even though the peak
circular velocity drops from, the shape of the circular velocity
profile does not change dramatically.  In particular, the central
circular velocity profile is still steeply rising.  
\citet{Stoehr:06} claim that the circular velocity profiles of 
subhaloes are best fit by a parabolic function, which is shallower 
than NFW circular velocity profiles, and that this confirms their earlier 
result that the internal structure and kinematics of the Milky 
Way satellites are in good agreement with the subhaloes found in 
CDM simulations \citep{Stoehr.etal:02}.
The persistence of the steep inner density structure in our 
satellite simulations conflicts with their assumed parabolic velocity profile. 
Recent very high resolution simulations following the formation of CDM haloes
in a cosmological context also have 
steep central density profiles in their subhaloes, which confirms our result
\citep{Diemand.etal:08,Springel.etal:08}.

\section{Estimating satellite mass loss}
\label{sec:massloss}

To estimate satellite mass loss, researchers often use the impulse
approximation.  The impulse approximation assumes that the
perturbation time scale is much shorter than the internal dynamical
time scale \citep{Spitzer87}.  However, in \S\ref{sec:circ} and
\S\ref{sec:ecc}, we demonstrated that satellite halo evolution results
from resonant effects, internal structure evolution, and tidal
truncation in addition to impulses.  In this section we will include
the two resonant effects described in \S\ref{sec:circ} to produce an improved
analytic estimate.  We implement these approximations in a simple
model that computes the mass loss by tidal truncation, gravitational
shocks, and resonant torques in spherical shells.  For the tests
presented here, we use the same satellite models and orbits presented
in \S\S\ref{sec:method}--\ref{sec:ecc}.  At each time step, we proceed
as follows:
\begin{enumerate}
\item The tidal radius $x_e$ is computed for the current position in
  the satellite's orbit.  The instantaneous satellite mass is the mass
  of all particles with $r<x_e$ and the remainder is considered to be
  stripped.
\item The satellite's mass \emph{outside} of the tidal radius is
  gradually stripped 
  on a crossing time scale.  In particular, for each time step we remove a
  mass fraction of $\Delta t/t_{orb}$, where $t_{orb}$ is the orbital period
  at the half mass radius of the satellite. This procedure prevents 
  unrealistic immediate and permanent satellite mass stripping outside the
  tidal radius (see \S\ref{sec:peri}).
\item The changes in energy and angular momentum for a given mass shell
  caused by the time-dependent external potential are computed in two parts:
  1) the change in energy from gravitational shocks is only computed 
  at every pericentre; 
  and 2) the change in angular momentum owing to the resonant torque 
  is computed at every step.
\item An increase in energy and angular momentum generally results in
  reducing the density of a mass shell. Overall, the reduction in
  density results in mass loss, decreases 
  $x_{e}$, and hence drives further mass loss.
\end{enumerate}
This mass loss algorithm is similar to the scheme used in
\citet{TB01,TB04,TB05a,TB05b}.  It provides the satellite mass loss
history, which can be used to compare with the simulation results.

We determine the tidal radius as the satellite radius where
$d\Phi_{eff} =0$ (see \S\ref{sec:method} and Appendix
\ref{sec:pot_eff}). Since this model does not include any velocity
information about individual satellite particles, our estimate cannot
include the Coriolis force. Instead we choose $\alpha=1.0$ when
evaluating $\Phi_{eff}$ to compensate for ignoring the Coriolis force.

For the gravitational shock heating, we compute the change in energy
using the impulse approximation with an adiabatic correction.
Assuming an spherical isothermal host halo, \citet{GHO99} derive the
following expression for the energy change:
\begin{eqnarray}
  \Delta E = \frac{1}{6}\left(\frac{\pi
      r}{R_{p}^{2}V_{p}}\right)^{2}\frac{M_{halo}}{R_{p}}A(x)
  \label{eq:tidal_E}
\end{eqnarray}
where $r$ is the satellite's radius, $R_{p}$ is the pericentre distance,
and $V_{p}$ is the satellite's velocity at pericentre.  $M_{halo}$
is the mass of the perturber, which for us is the enclosed halo mass at 
pericentre.  The quantity $A(x)$, the \emph{adiabatic}
correction, reduces the heating if the internal orbital time is
significantly shorter than the impulsive time scale.  We explore both
the Spitzer correction and the Weinberg correction
\citep{Weinberg94a,Weinberg94b,GO99}:
\begin{eqnarray}
  A(x) & = & \exp(-2x^{2}) \hspace{0.35\columnwidth}  \hbox{(Spitzer)}  \\
  A(x) & = & (1+x)^{-1.5} \hspace{0.35\columnwidth}  \hbox{(Weinberg)}
  \label{eq:addcorr}
\end{eqnarray} 
where $x = \omega\pi R_{p}/V_{p}$ and $\omega$ is the azimuthal
frequency of an orbit at $r$.  Equations
(\ref{eq:tidal_E})--(\ref{eq:addcorr}) determine the energy change for
each mass shell and for each shock event.  The added energy expands
the satellite.  Using the virial theorem, the expansion rate can be
estimated as $\Delta r = \Delta E r^{2}$ \citep{TB01}.  This satellite
expansion reduces the density in each shell: $\Delta \rho =
-(9/2\pi)(\Delta E/r^{2})$. We use this relationship to compute the
mass loss in Step (iv) of the algorithm above.

In addition to heating by gravitational shocks, the results from
\S\ref{sec:circ} confirm that heating by resonant torques plays an
important role in satellite mass loss.  This effect has not been
included in most studies of satellite evolution.  To accurately
estimate the resonant torque, one needs high quality N-body
simulations \citep[see][]{WK07a} or perturbation theory calculations
\citep{Weinberg86,Weinberg89}.  However, we can estimate the torque
using a simple model as follows.  First, we compute the torque on a
mass shell by assigning the mass in the shell to an equivalent ring.
The torque on the ring is computed using a simple spin-orbit coupling
calculation as in a planet--satellite interaction.  In this
approximation, we assume that the host halo potential is a point mass
at the centre of the host halo.  We model the quadruple moment of a
satellite mass shell as two equal, diametrically opposed point masses
in the satellite's orbital plane, separated by the diameter of the
mass shell.  Owing to the separation, the gravitational forces on the
two masses differ. The differential gravitational force torques the
two masses and we evaluate the total torque on the satellite's mass
shell by integrating this torque through the ring.

The mathematical form of the torque on the mass shell is then
\begin{eqnarray}
  \tau_{ring}(R) &=& 6GM\frac{R}{R_{sat}}
  \label{eq:torqueR}
\end{eqnarray} 
where $R_{sat}$ is the distance from the host halo centre to the
satellite, $R$ is the distance from the satellite centre to a given
mass shell, and $M$ is the enclosed host halo mass \citep[][Chapter
5.3 see their Fig. 5.7]{MD99}.  Equation (\ref{eq:torqueR}) provides
the specific torque on the satellite mass shell.

Second, we compute the fraction of resonant orbits in each mass shell.
Equation (\ref{eq:torqueR}) computes the torque on a rigid mass
shell, not the torque on the mass shell owing to the resonant
interaction.  To estimate the resonant torque, we estimate the mass
fraction of resonantly coupled orbits in the mass shell and multiply
the torque on the rigid mass shell by this fraction.  As shown in
equation (\ref{eq:res}), determining resonant orbits requires detailed
calculations.  
Using equation (24) of \citet{Weinberg94b}, \citet{GO99} show that 
the number of stars at the peak amplitude scales as $1/\tau$, where $\tau$ 
is the characteristic duration of the shock. The resonant shock and the 
resonant torque are based on the same physics with a different coupling and
frequency. By dimensional analysis, $1/\tau \sim \omega$, 
where $\omega$ is the angular frequency of satellite mass shell 
\footnote {The angular frequency $\omega = \frac{V_{c}}{R}$ where $V_{c}$ 
is the circular velocity and $R$ is the radius of a given mass shell.}.
Therefore, we estimate this fraction as the ratio of the azimuthal 
angular frequency of the orbiting satellite to the angular frequency 
of the satellite's mass shell.
Consequently, the torque on the satellite mass shell becomes:
\begin{equation}
  \tau_{sat}(R) = \tau_{ring}(R) \times
  \cases{
    \displaystyle{\left({\Omega_s\over\omega(R)}\right)} & 
    if $\Omega_{s} \le \omega (R)$ \cr
    \noalign{\medskip}
    \displaystyle{\left({\omega(R)\over\Omega_s}\right)} & 
    if $\Omega_{s} > \omega (R)$
  }
  \label{eq:torqueS}
\end{equation}
This torque leads to expansion and the expansion rate can
be estimated as
\begin{equation}
\Delta r = \sqrt{\frac{4r}{M_{sat}(r)}} \tau_{sat} \Delta t.
\end{equation}
This expansion reduces the satellite's density as $\Delta \rho \propto
-\frac{\Delta r}{r^{4}}$.  We also use this relationship to compute
the mass loss in Step (iv) of the algorithm above.

\begin{figure}
  \centerline{\includegraphics[width=1.0\columnwidth,angle=0]
    {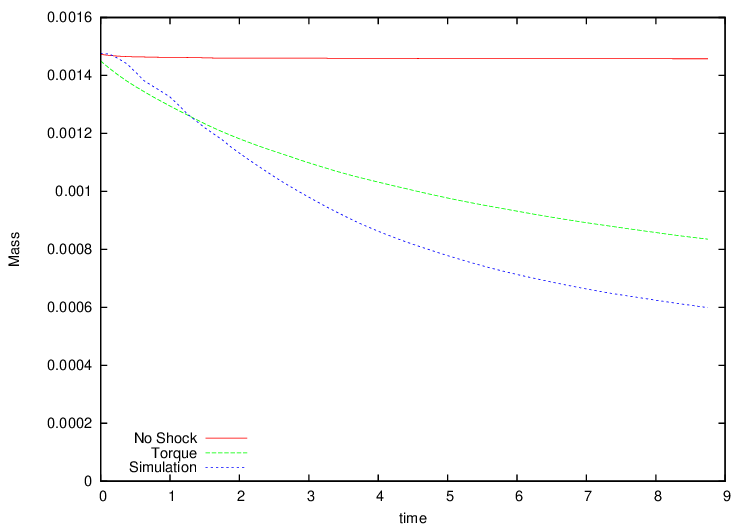}}
  \caption{The mass loss history of the circular orbit simulation.
    Analytic estimates of the mass loss history are compared with the
    simulation.  \emph{No Shock} is an analytic estimate only
    including tidal truncation and \emph{Torque} is an
    analytic estimate including the effects of tidal truncation and
    resonant torques.}
  \label{fig:massloss.XC1}
\end{figure}

\begin{figure}
  \centerline{\includegraphics[width=1.0\columnwidth,angle=0]
    {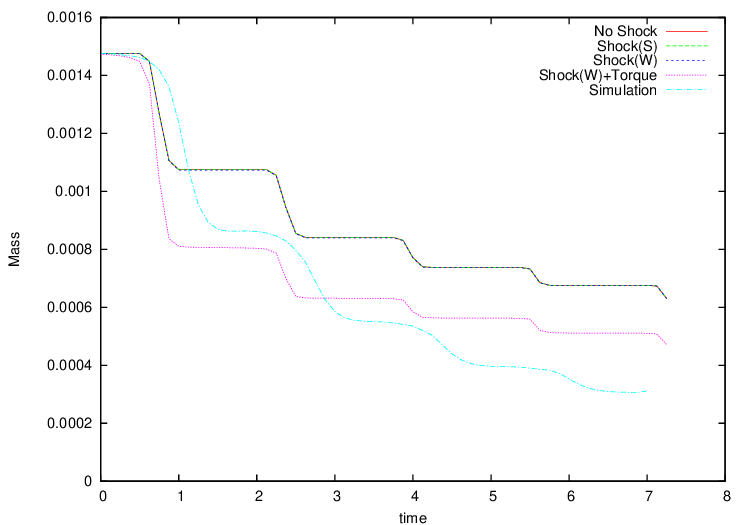}}
  \caption{The mass loss history for the eccentric orbit simulation.
    Analytic estimates of the mass loss history are compared with the
    simulation results.  \emph{No Shock} is an analytic estimate
    only including tidal truncation.  \emph{Shock(S)} is
    an analytic estimate including tidal truncation
    and impulse shock heating with the Spitzer correction.
    \emph{Shock(W)} is an analytic estimation including the effects of
    tidal truncation and impulse shock heating with the Weinberg
    correction.  \emph{Shock(W) + Torque} is the same as
    \emph{Shock(W)} but with the addition of the resonant torque.}
  \label{fig:massloss.XE1}
\end{figure}

\begin{figure}
  \centerline{\includegraphics[width=1.0\columnwidth,angle=0]
    {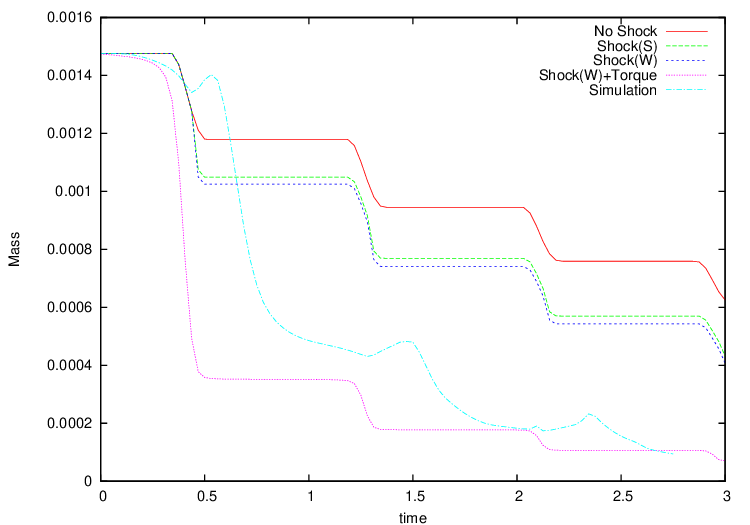}}
  \caption { The same as Fig. \ref{fig:massloss.XE1} but for the
    inner orbit simulation.  }
  \label{fig:massloss.XK2}
\end{figure}

The mass loss histories for the circular orbit simulation, the
eccentric orbit simulation, and the inner orbit simulation are
compared with the results of our mass-loss algorithm in Figs.
\ref{fig:massloss.XC1} -- \ref{fig:massloss.XK2}, respectively.  For
the circular orbit case, there are no gravitational shocks.  Fig.
\ref{fig:massloss.XC1} shows two analytic estimates: \emph{No Shock}
and \emph{Torque}.  The \emph{No Shock} estimate only includes tidal
truncation and predicts negligible mass loss.  The \emph{Torque} estimate
includes both the resonant torque approximation and tidal truncation.
Its mass loss history is much more similar to the simulation: more than 50\% of
the original mass is lost.  This suggests that mass loss for a
satellite on a circular orbit mainly results from resonant torques
and our algorithm for resonant torque provides a dramatically improved
description.  

For the eccentric orbit and inner orbit simulation cases, the
gravitational shock plays an important role in driving mass loss.
Figs. \ref{fig:massloss.XE1} and \ref{fig:massloss.XK2} compares our
mass loss algorithm to the simulations.  Here, we include the \emph{No
  Shock} model, the impulse approximation with the Spitzer correction
[\emph{Shock(S)}], the impulse approximation with the Weinberg correction
[\emph{Shock(W)}], and the impulse approximation with Weinberg
correction together with our resonant torque approximation
[\emph{Shock(W)+Torque}].  All the estimates include tidal truncation.
The \emph{Shock(W)+Torque} model best represents the mass loss seen in the
simulation while the \emph{Shock(W)} with no resonant torque predicts less
mass loss than that seen.  The importance of the
resonant torque is most obvious for the inner orbit simulation.
Compared to the simulation, the estimate including the shock and the
resonant torque [\emph{Shock(W)+Torque}] does significantly better
estimating the mass loss in the simulation than that of the shock alone, which
significantly underpredicts the mass loss.  Tidal truncation alone
[\emph{No Shock}] shows the worst agreement.

Although our new algorithm provides a dramatically improved satellite
mass loss history, discrepancies remain owing to the many complicated
physical processes involved in satellite disruption.  First, the
compressive gravitational shock is a strongly non-linear,
time-dependent perturbation.  Although the adiabatic correction
includes some of this time dependence by including the work done at
the frequency peak, many channels of possible coupling are ignored.
Second, satellite heating by resonant interactions is too complicated
to be accurately represented by our simple parametrisation.  For
example, the adiabatic correction extends the coupling to lower energies
by noting that the resonant coupling will have a power law rather than
an exponential scaling with frequency.  However, for a particular
interaction, individual resonances may dominate the response; this is
analogous to the difference between a \emph{line} and
\emph{continuous} spectrum.  Third, a real satellite halo has a wide
range of orbits, from circular to radial.  The mass shell
scheme cannot accurately capture the dynamics of these different
orbits.  Moreover, the entire satellite structure readjusts its
evolution in the course of satellite disruption.  This readjustment
process is also not included.  These arguments suggest that although our
new algorithm provides improved satellite mass loss histories, high
quality simulations are still necessary for an accurate prediction of
mass loss.

\section{Summary and Conclusions}
\label{sec:summary}

Using high resolution simulations with cosmologically motivated
initial conditions, we investigate the physical processes responsible
for the evolution of satellite galaxies in their host halo.  We
identified and explored the following important physical mechanisms
that result in satellite galaxy disruption.  Our main results are as
follows:
\begin{enumerate}
\item Resonant mechanisms of two types play a key role in satellite
  disruption: the resonant shock and the resonant torque.
\item We studied satellites on circular orbits to isolate the effects
  of resonant torques.  The ILR-like resonance $l_1:l_2:l_3=-1:2:2$
  (see equation (\ref{eq:res})) dominates the torque resonance because all
  the $l=1$ resonances and the 0:2:2 (corotation-like) and the 1:2:2
  (OLR-like) resonances are located outside the satellite's tidal
  radius.
\item Some important resonances require more particles than are
  typically used in cosmological simulations.  For example, to accurately
  reproduce the heating from the -1:2:2 resonance, a satellite
  simulation needs more than $10^{5}$ particles within the virial
  radius for our expansion code and possibly more for codes affected
  by small-scale noise, e.g. a tree code.  Too few particles results
  in less mass loss and it suggests that the lifetime of dark matter
  subhaloes in current cosmological simulations could be
  \emph{overestimated}.
\item For satellites on eccentric orbits, gravitational shocks
  dominate the heating but the heating rate is significantly
  underestimated by the impulse approximation for several reasons.
  First, strong low-order resonances couple to the phase space in the
  `adiabatic' regime and the work done on the satellite drives
  structural evolution.  Second, the ongoing tidal truncation affects
  the dynamics of the escaping material and the satellite's size.  The
  interplay of these mechanisms leads to continuous mass loss and
  evolution.
\item The number of resonances in a satellite increases as the
  satellite's orbital frequency increases; the strength of each
  resonant interaction also increases.  This affects satellites on
  low-energy and low-angular momentum orbits which have small
  pericentres and high satellite azimuthal frequencies.
\item Satellite mass stripping is an outside-in process in energy
  space.  We discuss the morphological consequences for the
  stripping signatures of stellar and gaseous streams.
\item We present an improved algorithm to estimate satellite mass loss
  including both resonant shocks and torques.  We show that the
  mass-loss history computed using this algorithm reproduces the
  general features seen in the simulations.
\end{enumerate}

The satellite evolution using N-body simulation has been studied by a
number of authors recently
\citep[e.g.][]{Hayashi.etal:03,Kazantzidis.etal:04,
  Boylan-Kolchin.Ma:07}. Their approach differs ours in several ways.
We use a highly idealised simulation configuration to better discern
the details of the dynamical mechanisms, focussing on the detailed
physical processes affecting satellite evolution such as resonant
dynamics that have not been rigorously addressed elsewhere.  For
example, we equilibrated the initially tidally truncated satellite to
reduce the artificial perturbation from sudden introduction of the
host halo potential.  We also explored a circular orbit simulation to
clearly demonstrate the effects of resonant torque.  We performed a
linear perturbation calculation to estimate the required resolution
for resonant effect, and made the simulations satisfy this
requirement. In addition, we use an expansion code to reduce the force
fluctuations on small scales. Owing to these efforts, we were able to
demonstrate the importance of the resonant torque in satellite
disruption.
  
\citet{Hayashi.etal:03} used an NFW halo as a satellite initial
conditions and a Tree code for a potential solver. They found that the
simple tidal-limit approximation underestimates the mass loss, as we
do, and found structural evolution, also similar to our findings.
\citet{Kazantzidis.etal:04} and \citet{Boylan-Kolchin.Ma:07} also
found that similar internal structure evolution using tree-code
simulations. This consensus suggests that the inner cusp of satellite
halo is not strongly affected by the tides from the host
halo. However, owing to their rather complicated configuration, they
could not discern the effect of the resonant torque, although some
authors \citep[e.g.][] {Hayashi.etal:03} has noticed that analytic
formulae underestimate the mass loss.

Although we have improved our understanding of the detailed physical
processes responsible for satellite disruption, some issues
remain.  By separating the heating mechanisms into two distinct
regimes, we achieved an improved understanding of resonant dynamics
for an eccentric orbit.  However, we have not compared this
approximation with a comprehensive perturbation theory
calculation. The resonant heating of satellites on eccentric orbits is
similar to heating by other subhaloes.  This interaction is an
important source of satellite evolution in addition to the interaction
with the smooth host halo.  In addition, the initially spherical
satellite is deformed during disruption owing to the host halo's tidal
field and we have not yet accounted for this deformation.  A
comprehensive treatment of this deformation might be necessary to
understand satellite disruption in detail.  Lastly, we also need a
better understanding of the non-linear processes that occur during
satellite evolution.

In this study, we characterised the linear processes; understanding
the detailed consequences of the non-linear processes is a daunting
future task.
We should then finally be able to fully constrain the satellite
disruption mechanism, which is an essential ingredient of galaxy
formation and evolution.

\section*{Acknowledgments}

This work was supported in part by NASA awards ATP NAGS-13308 and NAG5-12038.

\appendix
\section{A satellite's effective potential in a general gravitational
  field and a uniformly rotating frame}
\label{sec:pot_eff}

The effective potential of a spherical satellite as function of a
satellite's radius ($\bmath{r}$) is determined by its self-gravity, the
external potential, and the centrifugal force. We set $\bmath{r} =
\bmath{R} - \bmath{R_{0}}$ where $\bmath{R}$ is the location relative to the
host halo's centre and $\bmath{R_{0}}$ is the location of the
satellite's centre relative to the host halo's centre. The
acceleration in the satellite's frame, $\ddot{\bmath{r}}$, is difference
between the acceleration, $\ddot{\bmath{R}}$,
and the effective acceleration of the satellite, $\ddot{\bmath{R_{0}}}$,
in the host frame.  The quantities
$\ddot{\bmath{R}}$ and $\ddot{\bmath{R_{0}}}$ are
\begin{eqnarray}
 \ddot{\bmath{R}} & = & -\nabla \Phi_{tot}(\bmath{R}) - \bmath{\Omega} \times 
 (\bmath{\Omega} \times \bmath{R}), \\
  \ddot{\bmath{R_{0}}} & = & -\nabla \Phi_{tot}(\bmath{R_{0}}) - \bmath{\Omega} \times
 (\bmath{\Omega} \times \bmath{R_{0}}),
  \label{eq:R_Pot}
\end{eqnarray}
where $\Phi_{tot}(\bmath{R})$ is the total potential on the particle and
$\bmath{\Omega}$ is angular velocity of the satellite.  The quantity
$\Phi_{tot}(\bmath{R}) = \Phi_{host}(\bmath{R}) + \Phi_{sat}(\bmath{R})$,
where $\Phi_{host}(\bmath{R})$ is the host halo potential and
$\Phi_{sat} (\bmath{R})$ is the potential of the satellite.  Using
$\ddot{\bmath{r}} = \ddot{\bmath{R}} - \ddot{\bmath{R_{0}}}$, the equation
of motion in the satellite frame becomes
\begin{eqnarray}
 \ddot{\bmath{r}} & = & -\nabla \Phi_{sat}(\bmath{r}) -\nabla \Phi_{host}(\bmath{R}) \nonumber \\ & & 
 -\nabla \Phi_{host}(\bmath{R_{0}}) - \bmath{\Omega} \times (\bmath{\Omega} \times \bmath{r}).
  \label{eq:r_Pot}
\end{eqnarray}

Because we assume a circular orbit, $\Phi_{sat}(\bmath{R_{0}})$ is a constant;
$\nabla \Phi_{sat}(\bmath{R}) 
\rightarrow \nabla \Phi_{sat}(\bmath{r})$ and $\nabla \Phi_{tot} (\bmath{R_{0}})
\rightarrow  \nabla \Phi_{host}(\bmath{R_{0}})$. The second and third terms 
in equation (\ref{eq:r_Pot}) can be further simplified using a Taylor
expansion,
\begin{eqnarray}
  & - &\nabla \Phi_{host}(\bmath{R}) -\nabla \Phi_{host}(\bmath{R_{0}}) = \nonumber \\
  & - &\frac{d^{2} \Phi_{host}(\bmath{R_{0}})}{dR^{2}}\bmath{r} + {\cal
    O}\left[(r/R_o)^2\right].
 \label{eq:tayolr}
\end{eqnarray}
The last term in equation (\ref{eq:r_Pot}) is the centrifugal term.
Assuming that the orbital plane is equatorial with
${\hat\Omega}={\hat z}$ and $\bmath{r} = (x,y,x)$, we have
\begin{equation}
  - \bmath{\Omega} \times (\bmath{\Omega} \times \bmath{r}) = -[(\bmath{\Omega}
  \cdot \bmath{r})\bmath{\Omega} - (\bmath{\Omega}\cdot\bmath{\Omega})\bmath{r}] 
  =  \Omega^{2}(x,y,0).
  \label{eq:centri0}
\end{equation}
For generating initial conditions in \S\ref{sec:method} and our
idealised models in \S\ref{sec:massloss}, we will approximate equation (\ref{eq:centri0}) 
with a spherical average
\begin{equation}
  - \bmath{\Omega} \times (\bmath{\Omega} \times \bmath{r}) \approx \alpha
  \Omega^{2} \bmath{r}
\end{equation}
where $\alpha$ is between $0$ and $1$.

Using this expansion, equation (\ref{eq:r_Pot}) becomes
\begin{eqnarray}
  \ddot{\bmath{r}} & = & -\nabla \Phi_{sat}(\bmath{r}) -\frac{d^{2}\Phi_{host}
    (\bmath{R_{0}})}{dR^{2}}\bmath{r} + \alpha \Omega^{2} \bmath{r} \nonumber \\
  & = & -\nabla \Phi_{sat}(\bmath{r}) - [4\pi G \rho_{host}(R_{0}) -
    (2+\alpha)\Omega^{2}]\bmath{r}.
  \label{eq:r_Pot2}
\end{eqnarray}
Finally, a satellite's effective potential as function
of $\bmath{r}$ using equation (\ref{eq:r_Pot2}) can be written as:
\begin{eqnarray}
  \Phi_{eff}(\bmath{r}) & = &\Phi_{sat}(\bmath{r}) \nonumber \\ 
 & + &  \left[4\pi G \rho_{host}(\bmath{R_{0}}) - 
    (2+\alpha)\Omega^{2}\right] \times \left(\frac{1}{2}r^{2}\right).
  \label{eq:eff_Pot}
\end{eqnarray}


\end{document}